\documentclass[5p,twocolumn,times]{elsarticle}

\usepackage{multicol}

\usepackage{natbib}
\biboptions{sort&compress}

\usepackage{graphicx}
\usepackage{epstopdf, epsfig}
\usepackage{amsmath}
\usepackage{mwe,tikz}\usepackage[percent]{overpic}
\usepackage{tabularx}
\usepackage{subcaption}
\usepackage{epstopdf}
\usepackage{tikz}
\usepackage{babel}[english]
\usetikzlibrary{shapes,arrows}
\usepackage{array}
\usetikzlibrary{shapes.geometric}
\newcolumntype{L}[1]{>{\raggedright\let\newline\\\arraybackslash\hspace{0pt}}m{#1}}
\newcolumntype{C}[1]{>{\centering\let\newline\\\arraybackslash\hspace{0pt}}m{#1}}
\newcolumntype{R}[1]{>{\raggedleft\let\newline\\\arraybackslash\hspace{0pt}}m{#1}}
\usepackage{enumerate}
\usepackage{dcolumn}
\usepackage{bm}
\usepackage{xcolor} 
\usepackage{ulem}
\usepackage[colorlinks=true,linkcolor=blue,citecolor=blue,urlcolor=blue]{hyperref}
\usepackage[per-mode=symbol]{siunitx}         
\usepackage{xspace}
\usepackage{booktabs}
\usepackage{multirow}
\usepackage{adjustbox}
\usepackage{titlesec}
\usepackage{relsize}
\usepackage{colortbl}
\usepackage{steinmetz}
\usepackage{scalerel}
\usepackage{amssymb}
\usepackage{float}

\usepackage{lineno}   

\DeclareMathAlphabet{\mathcal}{OMS}{cmsy}{m}{n}
\newcommand{\bmp}[1]{\begin{minipage}{#1\columnwidth}}
\newcommand{\emp}{\end{minipage}}
\newcommand{\bea}{\begin{eqnarray}}
\newcommand{\eea}{\end{eqnarray}}

\newcommand{\s}{\,{\rm s}\xspace}

\newcommand{\ms}{\,{\rm m/s}\xspace}

\newcommand{\mm}{\,{\rm mm}\xspace}

\newcommand{\m}{\,{\rm m}\xspace}
\newcommand{\Hz}{\,{\rm Hz}\xspace}

\newcommand{\degree}{^{\circ}}

\newcommand{\RNum}[1]{\uppercase\expandafter{\romannumeral #1\relax}}

\usepackage{amssymb}
\usepackage{nomencl}
\usepackage{mdframed} 

\makenomenclature
\usepackage{etoolbox}
\renewcommand\nomgroup[1]{%
  \item[\bfseries
  \ifstrequal{#1}{A}{Abbreviations}{%
  \ifstrequal{#1}{L}{Letter Symbols}{%
  \ifstrequal{#1}{G}{Greek Symbols}{%
  \ifstrequal{#1}{S}{Subscripts and Superscripts}{%
  \ifstrequal{#1}{N}{Dimensionless Numbers}{}}}}}%
]}

\begin{document}

\begin{frontmatter}

    \title{Experimental Characterization of Non-Isothermal Sloshing in Microgravity}
    \author[1]{F. Monteiro\corref{cor1}%
    }
    \ead{francisco.monteiro@vki.ac.be}
    
    \author[1,2]{P. A. Marques}
    \author[1]{A. Simonini}
    \author[1]{L. Carbonnelle}
    \author[1]{M. A. M\'endez}
    \cortext[cor1]{Corresponding author}
    \address[1]{von Karman Institute for Fluid Dynamics, Waterloosesteenweg 72, Sint-Genesius-Rode, Belgium}
    \address[2]{Transferts, Interfaces Et Procédés (TIPs), Université Libre de Bruxelles, Av. Franklin Roosevelt 50, Brussels, 1050, Belgium}
    \date{\today}

\begin{abstract}
Sloshing of cryogenic liquid propellants can significantly impact a spacecraft's mission safety and performance by unpredictably altering the center of mass and producing large pressure fluctuations due to the increased heat and mass transfer within the tanks. This study, conducted as part of the NT-SPARGE (Non-isoThermal Sloshing PARabolic FliGht Experiment) project, provides a detailed experimental investigation of the thermodynamic evolution of a partially filled upright cylindrical tank undergoing non-isothermal sloshing in microgravity. Sloshing was induced by a step reduction in gravity during the 83$^{\text{rd}}$ European Space Agency (ESA) parabolic flight, resulting in a chaotic reorientation of the free surface under inertia-dominated conditions. To investigate the impact of heat and mass transfer on the sloshing dynamics, two identical test cells operating with a representative fluid, HFE-7000, in single-species were considered simultaneously. One cell was maintained in isothermal conditions, while the other started with initially thermally stratified conditions. Flow visualization, pressure, and temperature measurements were acquired for both cells. The results highlight the impact of thermal mixing on liquid dynamics coupled with the significant pressure and temperature fluctuations produced by the destratification. The comprehensive experimental data gathered provide a unique opportunity to validate numerical simulations and simplified models for non-isothermal sloshing in microgravity, thus contributing to improved cryogenic fluid management technologies.
\end{abstract}

    
\end{frontmatter}


\section{Introduction} \label{sec:introduction}

Recent advances in material science and rocket propulsion, combined with increased competition and interest in space exploration from nations and private entities, are propelling a new space race. Reusable and partially reusable launch vehicles and economies of scale are progressively reducing costs and democratizing access to space \cite{pekkanen_governing_2019}. Nevertheless, the development of deep space missions, space refueling, and orbital launch stations still require significant advances in propellant storage and management \cite{muratov_issues_2011, chato_low_2005, meyer_recent_2023,simonini_cryogenic_nature_2024}. Cryogenic propellants like liquid methane (LCH$_4$) and liquid hydrogen (LH$_2$) pose considerable challenges. These include losses from boil-off, the need for advanced thermal management systems, and issues with sloshing. Notably, these propellants can make up to \qty{90}{\percent} of a spacecraft's total mass \cite{holt_propellant_2009, isakowitz_international_2004}.

Liquid sloshing \cite{abramson_1981, ibrahim_2005} in partially filled tanks can significantly challenge the stability and maneuverability of spacecrafts \cite{dodge_2000}. Moreover, it can increase heat and mass transfer rates between liquid and ullage gasses, producing pressure fluctuations \cite{dreyer_propellant_2009} that could develop into structural instabilities or thrust oscillations \cite{werner_ballistic_2019}. Understanding, predicting, and controlling sloshing is particularly challenging because of its inherent nonlinear dynamics and its strong dependency on the forcing conditions, which greatly vary from hypergravity in the propulsive phase \cite{Ariane6_data} to microgravity in the ballistic phase \cite{behruzi_behavior_2006}.

Much of the early literature on sloshing dynamics has been driven by aerospace research during the 1960s and has focused on the dynamics and regime identification in isothermal conditions \cite{abramson_1981, ibrahim_2005, dodge_2000, royon-lebeaud, miles_1984, hopfinger_baumbach_2009}. Studies on non-isothermal effects have later targeted the impact on the tank's pressure evolution, particularly due to sloshing-induced evaporation or condensation \cite{moran_1994, lacapere_2009, ludwig_analysis_2014, arndt_2012, Vanforeest_2014, himeno_investigation_2009, himeno_heat_2011}. Nevertheless, all of the aforementioned studies considered on-ground (i.e., gravity-dominated) conditions. 

\begin{figure*}[t!]
\centering
\includegraphics[width=0.96\textwidth]{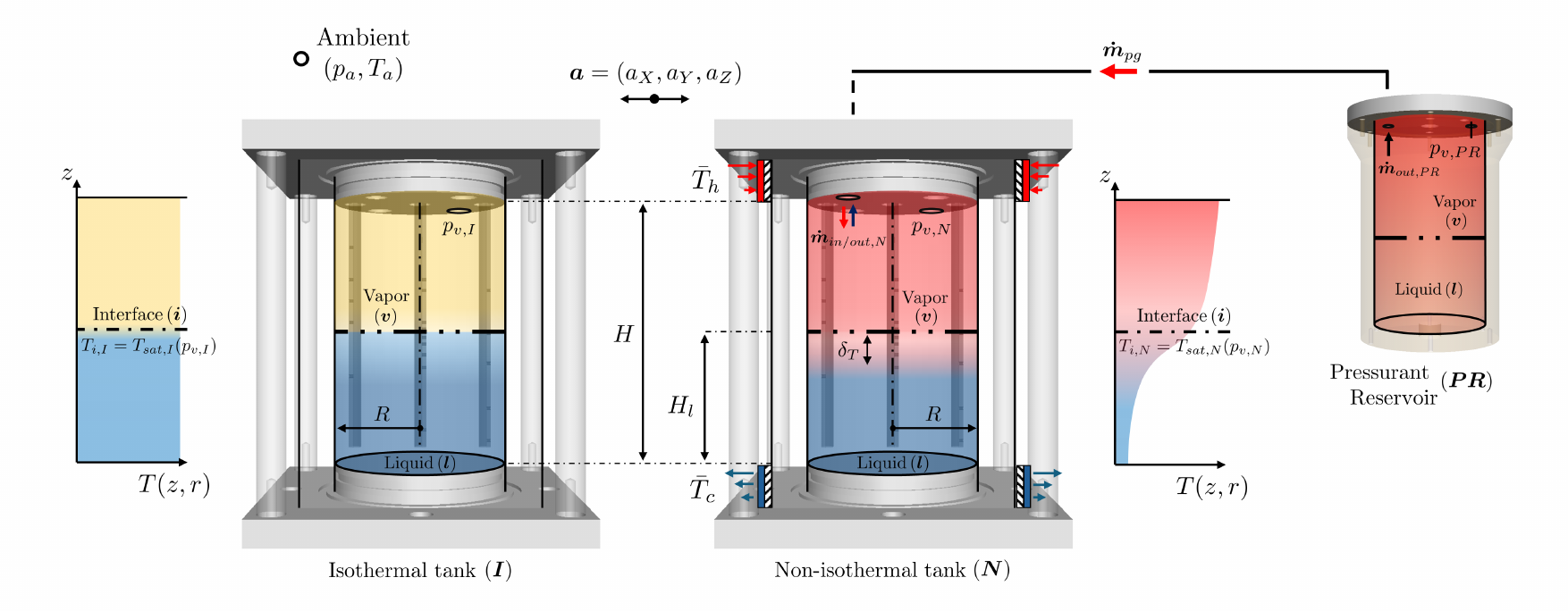}
\caption{Schematic of the NT-SPARGE concept. Two identical upright cylindrical tanks with radius $R$ and height $H$ were filled up to a level $H_l$. Cell $I$ is in isothermal conditions, while cell $N$ is in non-isothermal conditions, with the initial temperature profiles qualitatively sketched. In tank $N$, the stratification is set using a heater on the top at $\bar{T}_h$ and a cooler on the bottom at $\bar{T}_c$, together with the injection of warm vapor $\dot{m}_{pg}$ from an external reservoir $PR$. Both cells are then subjected to a step-like reduction of gravity, producing chaotic sloshing in inertia-dominated conditions. Pressure and temperature profiles are monitored and complemented with high-speed recording.} 
\label{fig:experimental_concept}
\end{figure*} 

The literature on non-isothermal sloshing in microgravity is mainly limited to the investigations under the COMPERE program \cite{dreyer_propellant_2009}, which have provided extensive databases from drop tower experiments \cite{schmitt_dreyer_2015,kulev_dreyer_2010, friese_liquid_2019}. Within this program, SOURCE-I and -II projects carried out on sounding rockets have focused on the dynamics of wetting, the heat and mass transfer near the contact line, and the possible occurrence of nucleate boiling in experiments \cite{fuhrmann_description_2008,fuhrmann_dreyer_2014, fuhrmann_free_2016}. 

Recently, the National Aeronautics and Space Administration (NASA) conducted a number of Zero Boil-Off Tank (ZBOT, \cite{kassemi_zero-boil-off_2018}) experiments within the Microgravity Science Glovebox (MSG) on the International Space Station (ISS). These experiments evaluated pressure control methods for managing non-isothermal environments within liquid propellant tanks in microgravity. The explored control technique aims to reduce liquid sub-cooling through axial jets, which improves temperature uniformity by promoting mixing \cite{kawanami_himeno_behavior_2019}. This pre-mixing decreases thermal destratification caused by sloshing, thus mitigating pressure fluctuations. The ZBOT studies are part of an extensive lineage of experimental campaigns initiated by the Tank Pressure Control Experiment (TPCE, \cite{bentz_tank_1997, bentz_tank_1990, hasan_tank_nodate}) carried out aboard the Space Shuttle, and followed by investigations on-ground with both experimental \cite{barsi_investigation_2009, barsi_investigation_2013} and numerical studies \cite{barsi_numerical_2008}. This research has provided valuable data on self-pressurization and pressure management under non-isothermal conditions, but only in scenarios where the liquid is quiescent or nearly quiescent. On the other hand, to the author's knowledge, all investigations on sloshing in microgravity have considered isothermal conditions (see \cite{simonini_microgravity,vergalla_experimental_2008, zhou_experimental_2010}), and no data could be found on sloshing-induced pressure fluctuations in microgravity conditions.

This work fills this gap by reporting on an extensive experimental characterization of non-isothermal sloshing in microgravity, focusing on the case of a cylindrical upright tank filled with HFE-7000 in equilibrium with its vapor and subject to a step-like reduction of gravity. The experiments were carried out on board the 83$^{\text{rd}}$ ESA parabolic flight campaign, reproducing a chaotic re-orientation of the gas-liquid interface. The initial conditions replicate the classic thermally stratified scenario one can expect at the end of a sufficiently long gravity-dominated phase \cite{Zhang2024}, with a superheated ullage vapor. This was accomplished using a dedicated experimental setup equipped with an active pressurization system. The setup allowed for high-speed flow visualization while simultaneously monitoring pressure and temperatures at multiple points. The sloshing dynamics in non-isothermal conditions are compared with an identical experiment in isothermal conditions. 

The article is structured as follows. Section \ref{sec:concept_goals} introduces the NT-SPARGE (Non-isoThermal Sloshing PARabolic FliGht Experiment) experiment's problem set and all the quantities of interest. Section \ref{sec:exp_setup} reports on the methodology, describing the experimental setup, including the isothermal and non-isothermal settings, the measurement chain, and measurement uncertainties. Section \ref{sec:exp_preparation} introduces the experiment preparation, including the degassing procedure to achieve a single-species environment. Section \ref{sec:exp_conditions} describes the experimental environment on board the 83$^{\text{rd}}$ ESA parabolic flight campaign, reporting on the measured accelerations. Section \ref{sec:results} presents the results in three subsections. The first focuses on setting the initial conditions for the sloshing experiments. The second reports on the thermal mixing due to the gravity step reduction for different initial fill ratios, while the third compares the dimensionless pressure response coupled to the acceleration profile variability. Conclusions and perspectives are presented in section \ref{sec:conclusions}.

\section{Concept and Goals of the NT-SPARGE Experiment} \label{sec:concept_goals}

The main components of the NT-SPARGE experiment are illustrated in Figure \ref{fig:experimental_concept}, together with the relevant quantities of interest. The reader is referred to \ref{sec:appA} for the nomenclature and list of symbols.

Two identical upright flat-bottom cylindrical tanks with radius $R$ and height $H$ are partially filled up to the same level $H_l$ with a liquid and maintained in an environment with ambient pressure $p_a$ and temperature $T_a$. Subscripts $v$ and $l$ are used to distinguish variables in the vapor and liquid phases, while the subscripts $N$ and $I$ refer to the different tanks: tank $N$ is in non-isothermal conditions (subsection \ref{subsec:exp_NIC}) while tank $I$ (subsection \ref{subsec:exp_IC}) is in isothermal conditions. Both tanks are single-species systems; hence the ullage volume $V_v$ is occupied only by the liquid's vapor at a pressure $p_{v,I}$ and $p_{v,N}$ and the gas-liquid interface is at the saturation temperature $T_{sat,I}(p_{v,I})$ and $T_{sat,N}(p_{v,N})$ (section \ref{sec:exp_preparation}). 

In tank $N$, the liquid bulk is subcooled at a temperature $T_{l,N}<T_{sat,N}$ while the vapor is superheated at a temperature $T_{v,N}>T_{sat,N}$. This condition is often encountered on tanks that had sufficiently long holding time; in our experiments, this is achieved by actively pressurizing the cell with pressurant gas (with flow rate $\dot{m}_{p g}$) at a temperature $T_{v, P R}>T_{v, N}$ from an external reservoir $PR$ (subsection \ref{subsec:exp_PR}) at a saturated vapor pressure $p_{v, P R}>$ $p_{v, N}$. The pressurization produces a sharp thermal gradient in the liquid beneath the interface characterized by a thermal boundary layer of thickness $\delta_T$ (subsection \ref{subsec:res1}). To further promote the thermal stratification and to control the experimental conditions, the top surface is heated at an average temperature $\bar{T}_h$; the bottom cover is cooled and maintained at a temperature $\bar{T}_c<\bar{T}_h$, setting the tank walls thermal profile $T_{w, N}(z, r)$. Figure \ref{fig:experimental_concept} sketches a qualitative temperature profile for each tank before the sloshing event. 


Both tanks are subjected to an acceleration $\boldsymbol{a}(t) = \left(a_X(t), a_Y(t), a_Z(t)\right)$, which in this work was produced by the airplane's parabolic trajectory (section \ref{sec:exp_conditions}). At time $t_{s, 0}=\SI{0.0}{\s}$, both tanks are under gravity-dominated conditions, with the liquid positioned in the lower part of the cells exhibiting a flat interface \cite{weislogel_mark_BOND, jang_mechanical_BOND}. The interface inclination is determined by the airplane's trajectory and the Bond number

\begin{equation}
\text{Bo}(t)=\frac{\rho_l(t) |\bm{a}(t)| R^2}{\sigma(t)}
\label{eq:bond}
\end{equation} is initially large ($\text{Bo} \gg 1.0 $).

A step reduction of vertical acceleration is initiated, bringing both tanks to microgravity conditions ($a_Z \sim 10^{-3} g_0$) with a chaotic reorientation of the liquid-free surface. This results in significant thermal mixing in the case of a non-isothermal environment, promoting phase change at the interface and eventually boiling. After a certain time in microgravity, the experiment returns to gravity-dominated conditions where a new thermodynamic equilibrium is achieved for the non-isothermal test cell. 

In this work, we focus on the pressure $p_{v, N}(t)$ and temperature $T_{N}(z,r,t)$ evolution under the impact of different fill ratios $H_l/R \in [0.0; 3.0]$. This controls the solid, gas, and liquid thermal stratification, and we assume that the thermal inertia of the solid is large enough to maintain the solid temperature constant within the time scale of the sloshing experiment. Although the solid properties can influence heat exchange with the fluid (see \cite{Ahizi2023, marques_assimilation_model}), we neglect these effects and treat the walls as adiabatic to focus exclusively on the scaling laws governing heat and mass transfer between the liquid and the ullage vapor.

We expect that the sloshing-induced mixing for a given acceleration profile $\boldsymbol{a}(t)$ to depend on the configuration geometry $\left(H, \, R\right)$, the initial conditions $\left(H_l(t_{s,0}), \, p_{v, N}(t_{s,0}), \, T_{N}(z,r,t_{s,0})\right)$ and the fluid properties: density $\rho$, kinematic viscosity $\nu$ and surface tension at the gas-liquid interface $\sigma$. We disregard the solid properties, which were kept equal in all experiments.

Considering $[u]\sim (R\,[a])^{1/2}$ as the reference velocity induced by the acceleration profile, with $[a]$ a reference scale for the external acceleration, one can relate the forcing conditions to the liquid properties and geometry through the Reynolds number ($\text{Re}$), the ratio between inertial and viscous forces: 

\begin{equation}
\text{Re}(t)=\frac{[a]^{1/2}R^{3/2}}{\nu_l(t)}.
\label{eq:reynolds}
\end{equation}

Moreover, we introduce two dimensionless numbers to scale the impact of liquid subcooling and vapor superheating on pressure evolution. Concerning the first ($\Pi_{l_0}$), assuming that the interface temperature $T_{i, N}(t_{s,0})$ sets the initial ullage pressure $p_{v, N}(t_{s,0})$ and at the limit of ideal mixing the interface drops to the liquid bulk temperature $T_{l, N}(t_{s,0})$, the coldest temperature in the system, following the Clausius–Clapeyron relation one could estimate a maximum pressure drop ratio 

\begin{equation}
\left(\frac{p_{sat,N}(T_l,t_{s,0})}{p_{v,N}(T_i,t_{s,0})}\right) \propto \exp \Biggl[\frac{\mathcal{L}_v}{\mathcal{R}_s}\Biggl(\frac{1}{T_{i,N}(t_{s,0})}-\frac{1}{T_{l,N}(t_{s,0})} \Biggr)\Biggr]=\Pi_{l_0},
\end{equation} where $\mathcal{L}_v$ is the specific latent heat of evaporation and $\mathcal{R}_s$ is the specific gas constant. Considering the oversimplification inherent to the Clausius-Clapeyron relation \cite{moran_shapiro}, $\Pi_{l_0}$ only provides an order of magnitude for the impact of liquid subcooling on the potential pressure drop.

Similarly, to scale and quantify the degree of superheating, we note that the ullage vapor at an initial temperature $T_{v, N}(t_{s,0})$ decreasing to $T_{l, N}(t_{s,0})$ could experience a maximum pressure drop ratio in the order of 

\begin{equation}
\left(\frac{p_{v,N}(t_{s,0})}{p_{sat,N}(T_l,t_{s,0})}\right) \propto \Biggl(\frac{T_{v,N}(t_{s,0})}{T_{l,N}(t_{s,0})}\Biggr)^{\frac{\gamma}{\gamma-1}}=\Pi_{v_0},
\end{equation} assuming an isentropic process, with $\gamma=c_p/c_v$ the heat capacity ratio. As for $\Pi_{v_0}$, this theoretical limit is unreachable due to heat transfer and irreversibilities. Nevertheless, this number provides an order of magnitude of how the superheated ullage gas could impact the pressure evolution. 

To summarize, considering the amplitude of the step-like reduction of gravitational acceleration as a scale $[a]$ and assuming inertia-dominated conditions, i.e., $\text{Re} \gg 1.0$, we might expect that the tank thermodynamic state is a function of the form: 

\begin{equation}
\hat{p}_{v,N}(\hat{t}), \hat{\Theta}(\hat{z},\hat{r},\hat{t})=f\left(\hat{t};\frac{H_l}{R}, \frac{H}{R},\Pi_{l_0},\Pi_{v_0},\text{Bo}(\hat{t}), \text{Re}(\hat{t}) \right).
\end{equation}

\begin{figure*}[ht!]
\centering
\includegraphics[width=0.81\textwidth]{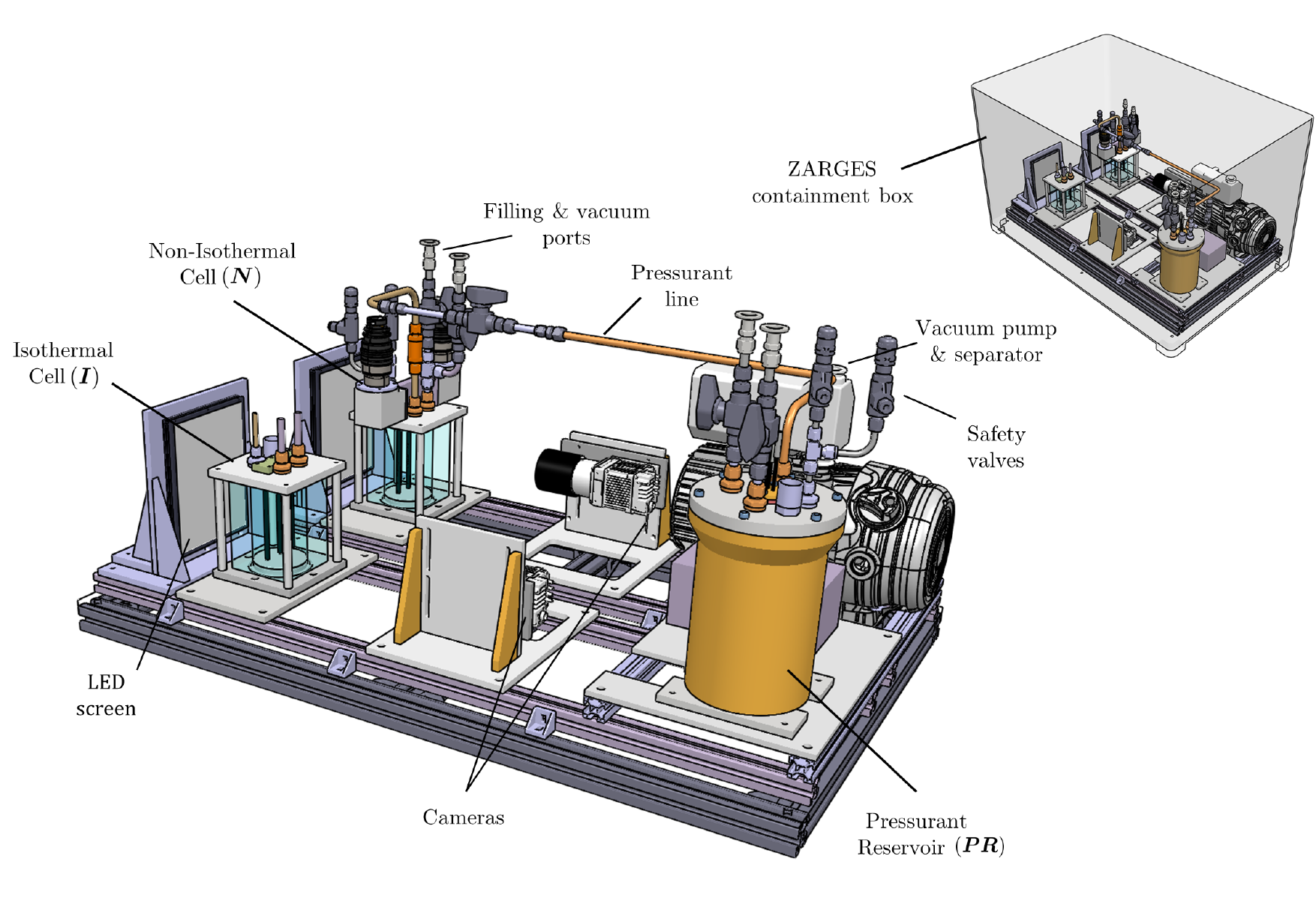}
\caption{NT-SPARGE setup integration in the 83$^{\text{rd}}$ ESA parabolic flight campaign. The experimental rack is composed of an isothermal quartz sloshing cell $I$, non-isothermal quartz sloshing cell $N$, a pressurant line with a throttling valve (TV), and a pressurant reservoir $PR$. The experiment containment system consists of a ZARGES box attached to the setup baseboard.} 
\label{fig:experimental_setup}
\end{figure*} 

\begin{table*}[t!]
\caption{Liquid saturation properties of HFE-7000 \cite{refprop, noauthor_3m_2022, rausch_properties}.}
\centering
\renewcommand{\arraystretch}{1.015} 
\begin{tabularx}{\textwidth}{X| *{8}{>{\centering\arraybackslash}X}} 
\toprule
$T$ [\unit{\K}] & $p$ [kPa] & $\rho$ [\unit{\kilo\gram\per\cubic\metre}] & $c_p$ [\unit{\kilo\joule\per\kilo\gram\K}] & $k$ [\unit{\watt\per\metre\per\kelvin}] & $\mathcal{L}_v$ [\unit{\kilo\joule\per\kilo\gram}] & $\sigma$ [\unit{\milli\newton\per\metre}] & $\mu$ [\unit{\micro\pascal\per\s}] & Pr [-] \\ 
\midrule[1.25pt] 
293    & 58.265  & 1418.3       & 1.2185            & 0.0758    & 139.66 & 12.314 & 512.16 & 8.251      \\
313     & 123.89   & 1362.5     & 1.2499            & 0.0699    & 131.38 & 10.265 & 392.74 & 7.033      \\
333     & 236.49    & 1302.6    & 1.2879            & 0.0645    & 122.41 & 8.2856 & 301.39 & 6.035      \\ 
343     & 316.12    & 1270.7    & 1.3102            & 0.0618    & 117.57  & 7.3252 & 263.61 & 5.598     \\ 
\bottomrule
\end{tabularx}
\label{tab:liquid_properties}
\end{table*}

\begin{table}[h!]
\caption{Vapor saturation properties of HFE-7000 (\cite{refprop, noauthor_3m_2022, rausch_properties}, molecular weight: $M_w = \SI{200.05}{\gram\per\mol}$ )}
\centering
\renewcommand{\arraystretch}{1.015} 
\begin{tabularx}{\columnwidth}{X| *{5}{>{\centering\arraybackslash}X}} 
\toprule
$T$ \small{[\unit{\K}]} & $p$ \small{[kPa]} & $\rho$ \small{[\unit{\kilo\gram\per\cubic\metre}]} & $c_v$ \small{[${\unit{\kilo\joule\per\kg\per\kelvin}}$]} & $k$ \small{[\unit{\watt\per\metre\per\kelvin}]} & $\gamma$ \small{[-]}  \\ 
\midrule[1.25pt] 
293     & 58.265    & 4.9798    & 0.8456         & 0.0109       & 1.062    \\
313     & 123.89    & 10.196    & 0.8875          & 0.0122      & 1.069  \\
333     & 236.49    & 19.042    & 0.9301          & 0.0135      & 1.079 \\ 
343     & 316.12    & 25.349    & 0.9517          & 0.0143      & 1.088  \\ 
\bottomrule
\end{tabularx}
\label{tab:gas_properties}
\end{table}

The dimensionless pressure evolution $\hat{p}_{v,N}(\hat{t})$ for the ullage volume is defined as:

\begin{equation}
\hat{p}_{v,N}(\hat{t})=\frac{p_{v,N}(\hat{t})-p_{sat,N}(T_l,t_{s,0})}{p_{v,N}(T_i, t_{s,0})-p_{sat, N}(T_l,t_{s,0})},
\label{eq:non_dim_pressure}
\end{equation} where the difference between $p_{v, N}(T_i, t_{s,0})$ and $p_{sat, N}(T_l,t_{s,0})$ sets the maximum possible ullage pressure drop due to the initial liquid subcooling condition at a temperature $T_{l, N}(t_{s,0})$. This dimensionless pressure equals $\hat{p}_{v, N}=1.0$ at sloshing onset and would reach $\hat{p}_{v, N}=0.0$ at the theoretical limit.

Similarly, the dimensionless temperature $\hat{\Theta}_{N}(\hat{z},\hat{r},\hat{t})$ is  

\begin{equation}
\hat{\Theta}_{N}(\hat{z},\hat{r},\hat{t})=\frac{T_{N}(\hat{z},\hat{r},\hat{t})-T_{i,N}(t_{s,0})}{T_{v, N}(t_{s,0})-T_{i,N}(t_{s,0})}
\label{eq:non_dim_temperature},
\end{equation} with $\hat{z}=z/H$ and $\hat{r}=r/R$ and the dimensionless time obtained by scaling $t$ with the initial heat diffusion's time scale $\hat{t}=t\alpha_v(t_{s,0})/R^2$. In the initial conditions, the temperature profile has $\hat{\Theta}_{N}=0.0$ at the interface $\hat{\Theta}_{N}<0.0$ in the liquid and $\hat{\Theta}_{N}>0.0$ in the gas. If the liquid temperature increases, the temperature profile becomes positive also in the liquid.

Within this framework, the NT-SPARGE project aimed to quantify how a sloshing event in microgravity can disrupt the temperature stratification in tank $N$, thereby causing large pressure fluctuations within a much shorter time scale than the time scale of thermal diffusion. Additionally, through the two test cells $N$ and $I$, subjected to identical acceleration profile $\boldsymbol{a}(t)$, we isolate the thermal effects from dynamic ones. This approach provides a unique experimental dataset for calibrating high-fidelity simulations and simplified thermodynamic models of non-isothermal sloshing (see for example \cite{grotle_2018, marques_assimilation_model, marques_experimental_2023}).       

\section{Experimental setup and measurement techniques} \label{sec:exp_setup} 

The main components of the experimental setup and its implementation into flight configuration on board the 83$^{\text{rd}}$ ESA parabolic flight campaign are detailed in Figure \ref{fig:experimental_setup}. Additional details on the heating and cooling elements, signal conditioners, acquisition, control, and powering systems for the two cells and active pressurization system are provided in subsections \ref{subsec:exp_NIC}, \ref{subsec:exp_IC} and \ref{subsec:exp_PR}. The camera acquisition setup, measurement chain, and uncertainties are covered in subsections \ref{subsec:exp_Camera} and \ref{subsec:exp_NIDAQ}.

The experimental rack is assembled into a containment structure with internal dimensions \qtyproduct{1175 x 793 x 717}{\mm}, to which a sealed cover (excluded for visualization purposes) is attached throughout the flight. This creates a closed environment with controlled pressure and temperature, monitored with an absolute pressure transducer AMS ME780 (Pressure $p_a$) and a type K welded tip fiberglass thermocouple (Temperature $T_a$) installed in the experiment base plate. A triaxial variable capacitance accelerometer (Endevco Model 7298-2) was attached to the experiment base plate between the sloshing cells. 

The experiments used 3M Novec 7000 Engineered Fluid (HFE-7000) \cite{noauthor_3m_2022}. This fluid is commonly used as a substitute for cryogenic propellants due to its low viscosity, surface tension, and extremely high volatility. Tables \ref{tab:liquid_properties} and \ref{tab:gas_properties} provide liquid and vapor saturation properties for HFE-7000 within the temperature range of the current experimental campaign. 

\subsection{Non-isothermal sloshing cell} \label{subsec:exp_NIC}

\begin{figure}[t!]
\centering
\includegraphics[width=0.95\columnwidth]{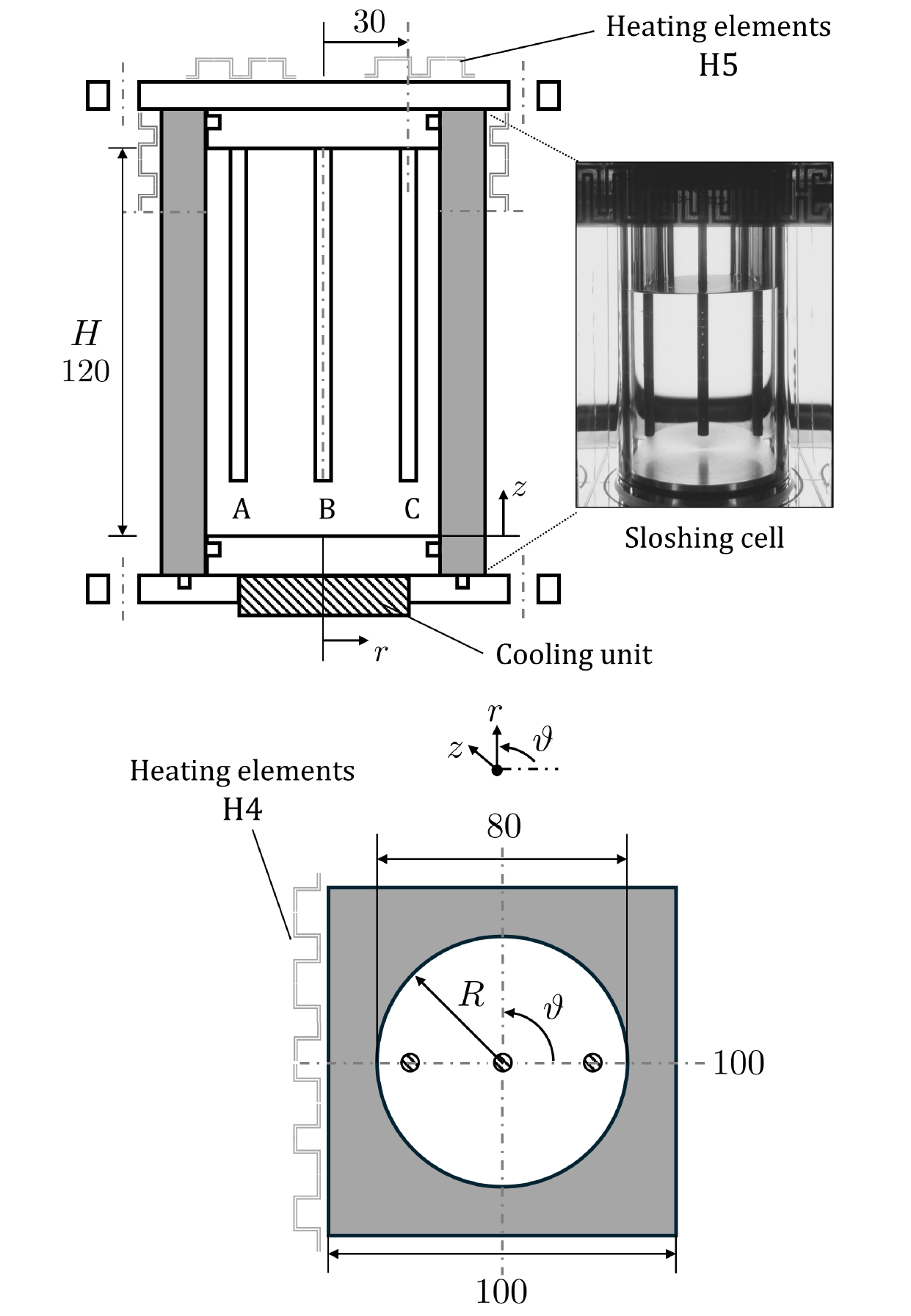}
\caption{Schematic of the non-isothermal sloshing cell. (Top): Cell $r-z$ plane view with the inner thermocouple racks position. A backlight visualization from the camera's point of view is also provided. (Bottom): Cell $r-\vartheta$ plane. All dimensions are in millimeters [mm].} 
\label{fig:cell}
\end{figure} 

The non-isothermal sloshing cell $N$ is illustrated in Figure \ref{fig:cell}. It consists of a rectangular quartz block (\qtyproduct{100 x 100 x 144}{\mm}) in which a passing-through hole of \SI{80}{\mm} diameter ($R = \SI{40}{mm}$) was drilled.
The cell was positioned between two covers, held in place by four steel rods attached to an aluminum base plate. The covers consist of \SI{11.5}{\mm} aluminum plates with a central cylindrical extrusion, \SI{12}{\mm} thick. Each cover features a double O-ring assembly: one ring aligns with the inner walls of the sloshing cell, while the upper and lower faces compress the second ring. The inner cell height is $H=\SI{120}{\mm}$, providing a volume of \SI{603.19}{\cubic\centi\metre}. The top cover includes ports for pressurization, filling, vacuum, and pressure safety release, contributing an additional vacant volume of \SI{15.48}{\cubic\centi\metre}. The volume for each component is reported in Table \ref{tab:volumes_cell}. 
A brass diffuser was added to the pressurant inlet line to minimize the pressurization jet impingement on the gas-liquid interface by redirecting the incoming superheated vapor radially toward the cell walls. Table \ref{tab:material_properties} provides the material properties for the quartz and aluminum.

\begin{table}[t!]
\caption{Void (a) and solid (b) volumes for the sloshing cells.}
\begin{tabularx}{\columnwidth}{X| *{3}{>{\centering\arraybackslash}X}} 
\toprule
Component           & Volume [cm$^3$] \\ 
\midrule[1pt] 
\multicolumn{2}{c}{(a) Void}       \\
\midrule[1.25pt] 
Quartz cell         & 603.19       \\
Filling port        & 5.18         \\
Vacuum port         & 2.99         \\
Safety valve no.01  & 1.83         \\
Safety valve no.02  & 2.04         \\
Pressurization port & 3.44         \\
\midrule[1pt] 
\multicolumn{2}{c}{(b) Solid}      \\
\midrule[1.25pt] 
Thermocouples rod   & 3.20         \\
Diffuser            & 0.59         \\ 
\bottomrule
\end{tabularx}
\label{tab:volumes_cell}
\end{table}

\begin{table}[h!]
\caption{Relevant material properties.}
\centering
\renewcommand{\arraystretch}{1.025} 
\begin{tabularx}{\columnwidth}{X| *{3}{>{\centering\arraybackslash}X}} 
\toprule
Material & $\rho$ [\unit{\kilo\gram\per\cubic\metre}] & $c_p$ [${\unit{\kilo\joule\per\kg\per\kelvin}}$] & $k$ [\unit{\watt\per\metre\per\kelvin}] \\ 
\midrule[1.25pt] 
Brass                   & 8500   & 0.380    & 109     \\
Aluminum           & 2700    & 0.900    & 205       \\ 
Quartz          & 2650  & 0.830    & 1.30      \\
Bakelite             & 1250    & 1.100    & 0.20        \\ 
Cork             & 240    & 1.200    & 0.04        \\ 
\bottomrule
\end{tabularx}
\label{tab:material_properties}
\end{table}

\begin{table}[t!]
\centering
\renewcommand{\arraystretch}{1.025} 
\caption{Position of the thermocouples in the sloshing cell for the rods A, B, C according to the cylindrical coordinates $(r,\vartheta,z)$.}
\begin{tabularx}{\columnwidth}{X| *{3}{>{\centering\arraybackslash}X}} 
\toprule
Reference   & $z$ [mm] & $r$ [mm] & $\vartheta$ [ºdeg.] \\
\midrule[1.25pt] 
Tc. $A_{5,N}$    & 100    & 33.50  & 180       \\
Tc. $A_{4,N}$    & 85     & 33.50  & 180       \\
Tc. $A_{3,N}$    & 75     & 33.50  & 180       \\
Tc. $A_{2,N}$    & 65     & 33.50  & 180       \\
Tc. $A_{1,N}$    & 40     & 33.50  & 180       \\
\midrule[0.5pt] 
Tc. $B_{10,N}$   & 120    & 3.50   & 270       \\
Tc. $B_{9,N}$    & 85     & 3.50   & 270       \\
Tc. $B_{8,N}$    & 80     & 3.50   & 270       \\
Tc. $B_{7,N}$    & 65     & 3.50   & 270       \\
Tc. $B_{6,N}$    & 62     & 3.50   & 270       \\
Tc. $B_{5,N}$    & 58     & 3.50   & 270       \\
Tc. $B_{4,N}$    & 55     & 3.50   & 270       \\
Tc. $B_{3,N}$    & 50     & 3.50   & 270       \\
Tc. $B_{2,N}$    & 44     & 3.50   & 270       \\
Tc. $B_{1,N}$    & 15     & 3.50   & 270       \\
\midrule[0.5pt] 
Tc. $C_{5,N}$    & 80     & 33.50  & 0         \\
Tc. $C_{4,N}$    & 60     & 33.50  & 0         \\
Tc. $C_{3,N}$    & 46     & 33.50  & 0         \\
Tc. $C_{2,N}$    & 25     & 33.50  & 0         \\
Tc. $C_{1,N}$    & 10     & 33.50  & 0         \\ 
\bottomrule
\end{tabularx}
\label{tab:Tcs_inside_cell}
\end{table}

\begin{table}[t!]
\caption{Sensor location in the top cover and external walls of the sloshing cell according to the cylindrical coordinates $(r,\vartheta,z)$.}
\centering
\renewcommand{\arraystretch}{1.025} 
\begin{tabularx}{\columnwidth}{X| *{3}{>{\centering\arraybackslash}X}} 
\toprule
Reference   & $z$ [mm] & $r$ [mm] & $\vartheta$ [ºdeg.] \\
\midrule[1.25pt] 
Tc. $D_{7,N}$      & 144    & 60     & 45        \\
Tc. $E_{3,N}$      & 144    & 50     & 315       \\
Tc. $D_{1,N}$      & 85     & 50     & 180       \\
Tc. $D_{2,N}$      & 65     & 50     & 180       \\
Tc. $D_{3,N}$      & 50     & 50     & 180       \\
Tc. $D_{4,N}$      & 40     & 50     & 180       \\
Tc. $D_{6,N}$      & 25     & 50     & 0         \\
Tc. $D_{5,N}$      & 10     & 50     & 0         \\
\midrule[0.5pt]
$p_{v,N}$ & 164    & 27     & 45       \\
\bottomrule
\end{tabularx}
\label{tab:Tcs_out_cell}
\end{table}

Twenty type K ultra-fine wire perfluoroalkoxy (PFA) insulated thermocouples from TC Direct were employed across three stainless steel rods. The junction diameter of \SI{0.08}{\mm} allows for a swift time response estimated to be below \SI{15}{\ms}. Two rods (Figure \ref{fig:cell}, A and C) are positioned near the walls to characterize the liquid rise due to capillary and inertial forces. Meanwhile, the central rod (Figure \ref{fig:cell}, B) retrieves the liquid thermal stratification before each parabola when the liquid is under normal gravity conditions and settles at the bottom of the tank. Each rod has a volume of approximately \SI{3.20}{\cubic\centi\metre}. Feedthrough connectors convey the thermocouple wires from inside the cell to the acquisition board. Due to manufacturing constraints, no sensor could be glued to the inner quartz walls.

The temperature on the outside quartz walls was measured using six thermocouples, while two thermocouples were installed on the top cover. These \SI{0.08}{\mm} type K thermocouples were glued on the solid wall with highly thermal conductive epoxy OMEGABOND 100. The detailed positions for the sloshing cell inner and outer sensors is given in Tables \ref{tab:Tcs_inside_cell} and \ref{tab:Tcs_out_cell} according to the reference in Figure \ref{fig:cell}. The pressure signal was acquired through a diaphragm (AMSYS ME780) assembled in a stainless-steel two-piece housing, flush mounted to the top cover to minimize the sensor response time (estimated at below \SI{5}{\ms}). 

Heating elements were added to the outside walls of the quartz cell (H4) and the aluminum top cover (H5) to set the temperature $\bar{T}_h$ (see Figure \ref{fig:experimental_concept}) and thermally stratify the solid. Heaters on the quartz consist of four MINCO polyimide thermofoils (\qtyproduct{101.6x25.4}{\mm}) glued to the upper part of the cell’s external walls (maximum power of \SI{66.5}{\watt} with a voltage of \SI{39}{\volt}). The two heaters on the aluminum top cover are flexible polyimide foils (\qtyproduct{45x100}{\mm}) with a maximum power of \SI{30}{\watt} at a voltage of \SI{24}{\volt}. Both were attached using self-adhesive material. A proportional integral derivative (PID) controlled power unit regulates the supplied power, receiving a temperature input from a thermocouple in the aluminum top cover. The PID setpoint value was settled at $\bar{T}_h = 333.0\pm\SI{2.5}{\kelvin}$ during the experiment. No insulation was used, ensuring visual access to the cell is only restricted by the quartz heating foils (subsection \ref{subsec:exp_Camera}).

Likewise, to ensure the thermal stratification over the whole duration of a parabolic flight (an average of 2 hours and 15 minutes), a cooling unit composed of a direct-to-liquid thermoelectric assembly (maximum cooling power of \SI{74}{\watt}) was installed on the cell's bottom cover. The heat is dissipated to a closed-loop liquid circuit. The cold-plate (\qtyproduct{100 x 60}{\mm}) target temperature is established via a \SI{0.3}{\mm} type K welded tip fiberglass temperature probe connected to the regulation system. The PID reference temperature for the bottom cover was defined as $\bar{T}_c = 293.0\pm\SI{2.5}{\kelvin}$.

\subsection{Isothermal sloshing cell} \label{subsec:exp_IC}

The isothermal sloshing cell $I$ is identical to the non-isothermal sloshing tank $N$. This subsystem was also equipped with temperature rods, pressurization, filling, vacuum, and safety ports in the top cover to reproduce the exact geometry of the $N$ cell. Nonetheless, no thermocouples were placed in the temperature rods. To characterize the thermodynamic environment in this cell, an AMSYS ME780 pressure transducer $p_{v, I}$ was employed, and a \SI{0.3}{\mm} type K welded-tip fiberglass thermocouple in the top cover was used as a temperature reference, assuming negligible thermal gradients. Table \ref{tab:ISO_CELL_intrumentation} summarizes the instrumentation position.

\begin{table}[h!]
\centering
\renewcommand{\arraystretch}{1.025} 
\caption{Sensor position at the isothermal sloshing cell top cover according to the cylindrical coordinates $(r,\vartheta,z)$.}
\begin{tabularx}{\columnwidth}{X| *{3}{>{\centering\arraybackslash}X}} 
\toprule
Reference   & $z$ [mm] & $r$ [mm] & $\vartheta$ [ºdeg.] \\
\midrule[1.25pt] 
Tc. $E_{4, I}$      & 164    & 30      & 45         \\
\midrule[0.50pt] 
$p_{v, I}$ & 164      & 27      & 45         \\
\bottomrule
\end{tabularx}
\label{tab:ISO_CELL_intrumentation}
\end{table} 

\subsection{Active-pressurization system operation} \label{subsec:exp_PR}

\begin{figure}[t!]
\centering
\includegraphics[width=0.95\columnwidth]{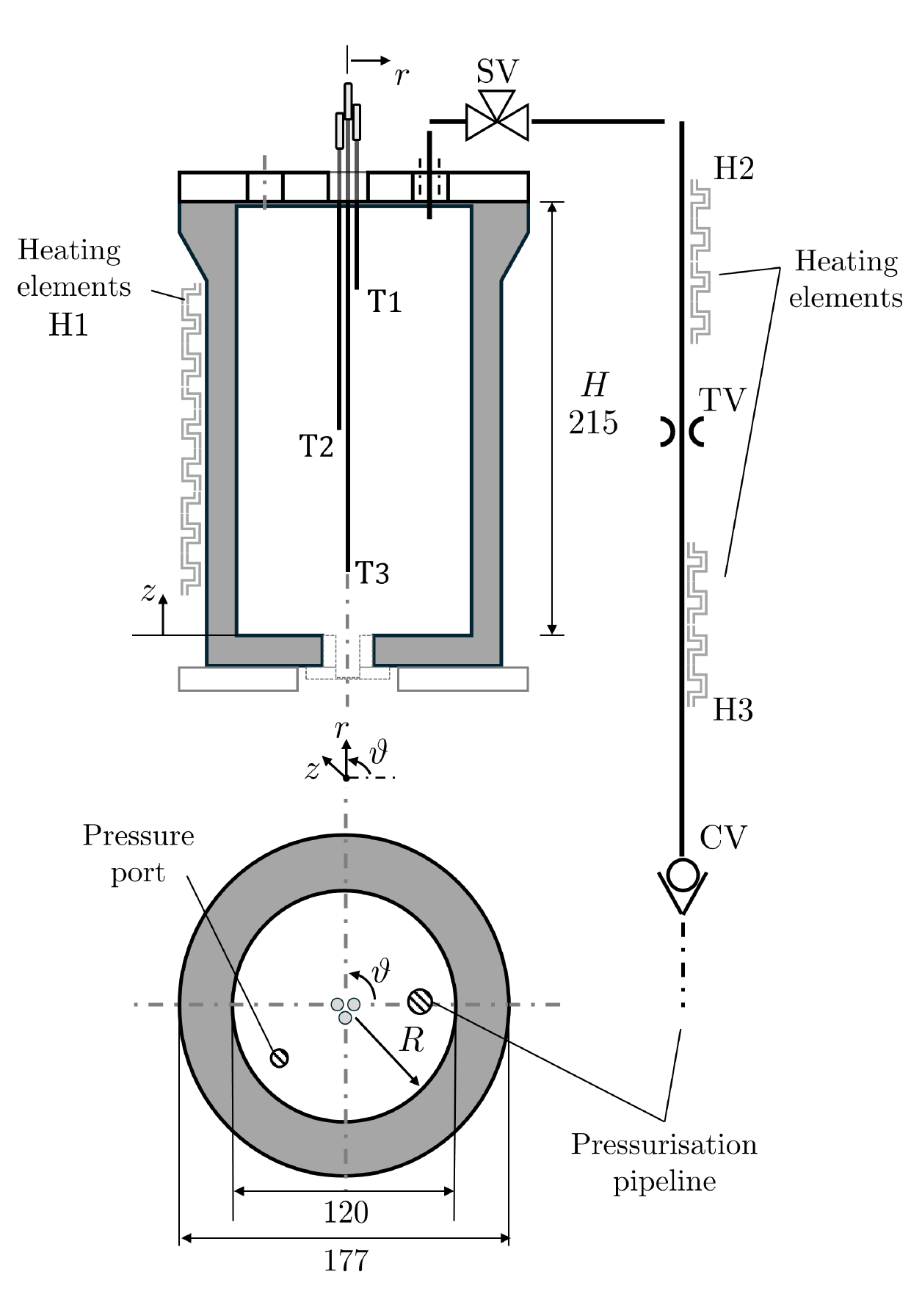}
\caption{Schematic representation of the active-pressurization system. (Top): Pressurant reservoir $r-z$ plane view with the position and assembly of the thermocouples sketched. (Right): Hydraulic and heating elements sketched and named along the pressurization pipeline. (Bottom): Reservoir $r-\vartheta$ view with radial dimensions. All dimensions are in millimeters [mm].} 
\label{fig:PR}
\end{figure}

The active pressurization system used to set the initial conditions in the non-isothermal quartz sloshing cell $N$ consists of a pressurant reservoir storing superheated gas and a pressurant line. A schematic of the pressurant reservoir $PR$ is provided in Figure \ref{fig:PR}. This is a brass cylinder with an outer height of \SI{230}{\mm} and an outer diameter of \SI{150}{\mm}. The inner height is \SI{215}{\mm}, and the inner diameter is \SI{120}{\mm}.

The tank is closed by a cylindrical top cover in aluminum (\SI{15}{\mm} of thickness). The reservoir volume is \SI{2431.59}{\cubic\centi\metre}. A bakelite plate supports the pressurant reservoir to minimize heat transfer to the test bench. At the bottom, a ball valve acts as an emptying port, with a circuit volume of \SI{8.25}{\cubic\centi\metre}. Additionally, the hydraulic element ports in the top cover with Klein Flange (KF) connectors for vacuuming, filling, and pressure release add up to a volume of \SI{10.03}{\cubic\centi\metre}. 

\begin{table}[t!]
\centering
\renewcommand{\arraystretch}{1.025} 
\caption{Thermocouples and pressure sensors position on the pressurant reservoir $PR$. The reference is located at the reservoir's bottom centerline according to the cylindrical coordinates $(r,\vartheta,z)$.}
\begin{tabularx}{\columnwidth}{X| *{3}{>{\centering\arraybackslash}X}} 
\toprule
Reference   & $z$ [mm] & $r$ [mm] & $\vartheta$ [\unit{\degree}] \\
\midrule[1.25pt] 
Tc. $T_{1, PR}$      & 172    & 5      & 0         \\
Tc. $T_{2, PR}$      & 100    & 5      & 180       \\
Tc. $T_{3, PR}$     & 15     & 5      & 270       \\
\midrule[0.5pt]
$p_{v, PR}$     & 250    & 47     & 225       \\
$p_{l, PR}$     & -20      & 80      & 0         \\
\bottomrule
\end{tabularx}
\label{tab:PRintrumentation}
\end{table}

A band heater (H1, Vulcanic mica) with a diameter of \SI{150}{\mm} and height of \SI{160}{\mm} surrounds the tank. This has a maximum power of \SI{2800}{\watt} at \SI{230}{\volt} and is coupled to a PID controller connected to a temperature sensor in the outer shell of the heating band. Two bi-metallic thermal switches opening at \SI{398.0}{\K} connected in series were fixed to the reservoir brass outer shell to ensure fault tolerance. This arrangement ensures a controlled thermodynamic condition while guaranteeing that the system autonomously returns to its predefined initial state after each pressurization cycle. The heater is externally insulated with a \SI{8}{\mm} thickness cork sheet with material properties provided in Table \ref{tab:material_properties}.

This reservoir includes three \SI{3.0}{mm} mineral-insulated type K thermocouples (TC Direct) with pot seals assembled through an O-ring compression setup. The thermocouples have a response time of \SI{800}{\ms}, and their positions are defined in Table \ref{tab:PRintrumentation}. Pressure signals were obtained via AMS ME780 transducers: one was flush-mounted to the reservoir top cover $p_{v, PR}$, and the other was connected to the bottom emptying port $p_{l, PR}$.

The pressurant line (Figure \ref{fig:PR}) transports the superheated vapor from the reservoir to the non-isothermal tank. It is a standard stainless steel pipe with an external diameter of \SI{9.525}{\mm} and \SI{1.651}{\mm} wall thickness. A 2-way On/Off brass direct action solenoid valve (SV) with an orifice diameter of \SI{4.5}{\mm} and internal threaded connection is placed at the reservoir outlet (enclosing the vapor). This valve is remotely operated when a pressurization cycle is performed. An upstream fixed-angle ball valve acts as a throttling device (TV), setting the pressurization conditions (mass flow rate $\dot{m}_{pg}$), and a swing-check valve (CV) is installed at the inlet of the sloshing cell. The line void volume between the solenoid valve and cell inlet is approximately \SI{39.94}{\cubic\centi\metre}. 

Two silicone rubber wire-wound heaters (MINCO) of size \qtyproduct{762 x 25.4}{\mm} are clamped to the line in a spiral-shaped assembly to minimize vapor energy losses during pressurization. One of these is placed between the solenoid and throttling valve (H2), and one is placed near the sloshing cell inlet (H3). The heater has a maximum power of \SI{50}{\watt} with a supply voltage of \SI{120}{\volt}. A fixed temperature boundary condition at the pipeline surface is imposed through a PID controller that receives a temperature signal from a \SI{0.3}{mm} type K welded tip fiberglass thermocouple placed between the heater and the pipeline's wall. The heating assembly and pipeline are externally insulated through an AP ArmaFlex foam tube with a thickness of \SI{25}{mm}. The $PR$ reservoir and the pipeline were kept at an external fixed temperature of approximately $\bar{T}_{PR} = 343.0\pm\SI{2.5}{\kelvin}$ throughout the experiment.

\subsection{High-speed video recording} \label{subsec:exp_Camera}

\begin{figure}[t!]
    \centering   
    \begin{subfigure}{0.45\columnwidth}
        \centering
        \includegraphics[width=1.0\textwidth]{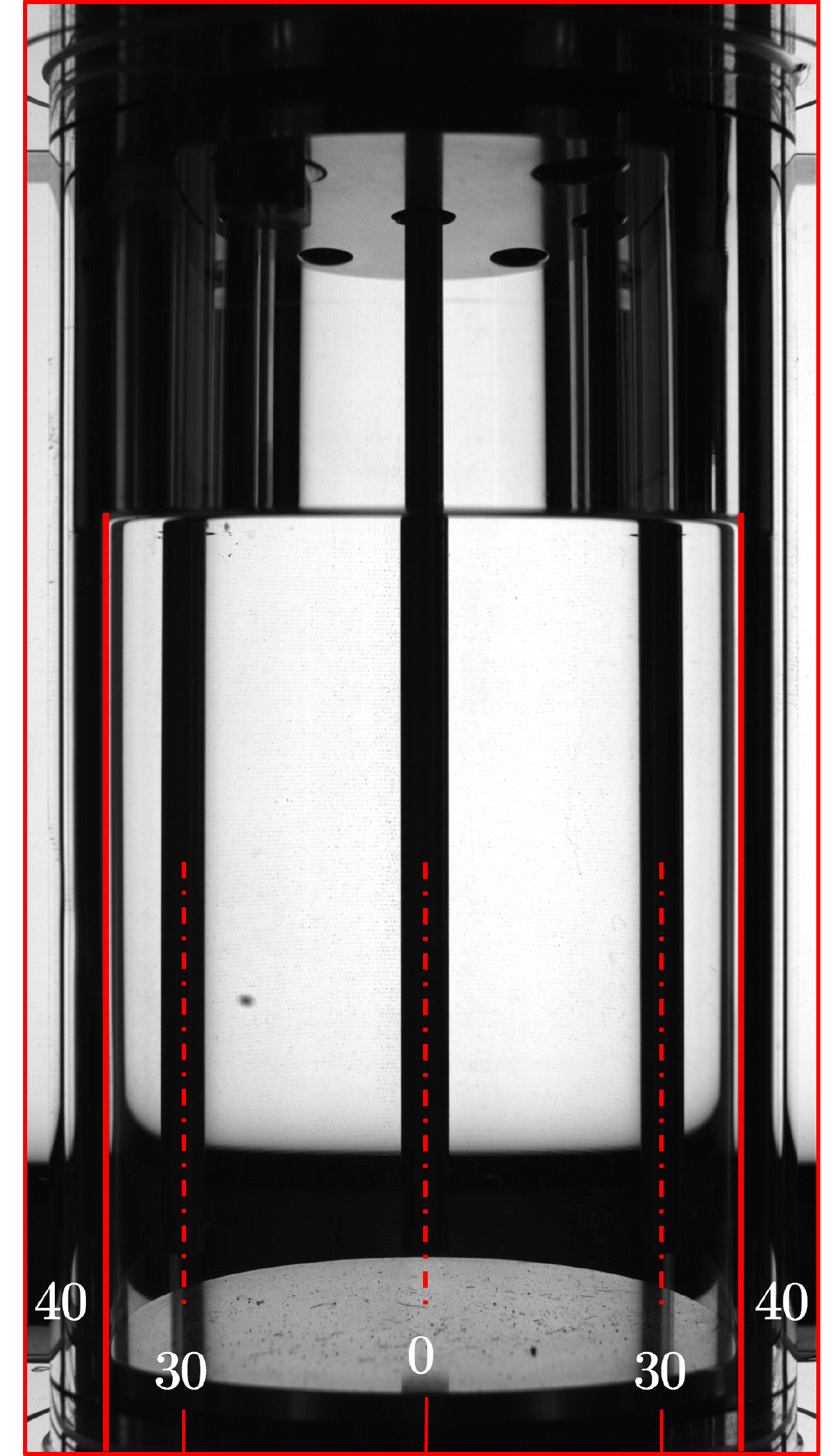}
        \caption{Isothermal sloshing cell $I$.}
    \end{subfigure}
    \hfill
    \begin{subfigure}{0.45\columnwidth}
        \centering
        \includegraphics[width=1.0\textwidth]{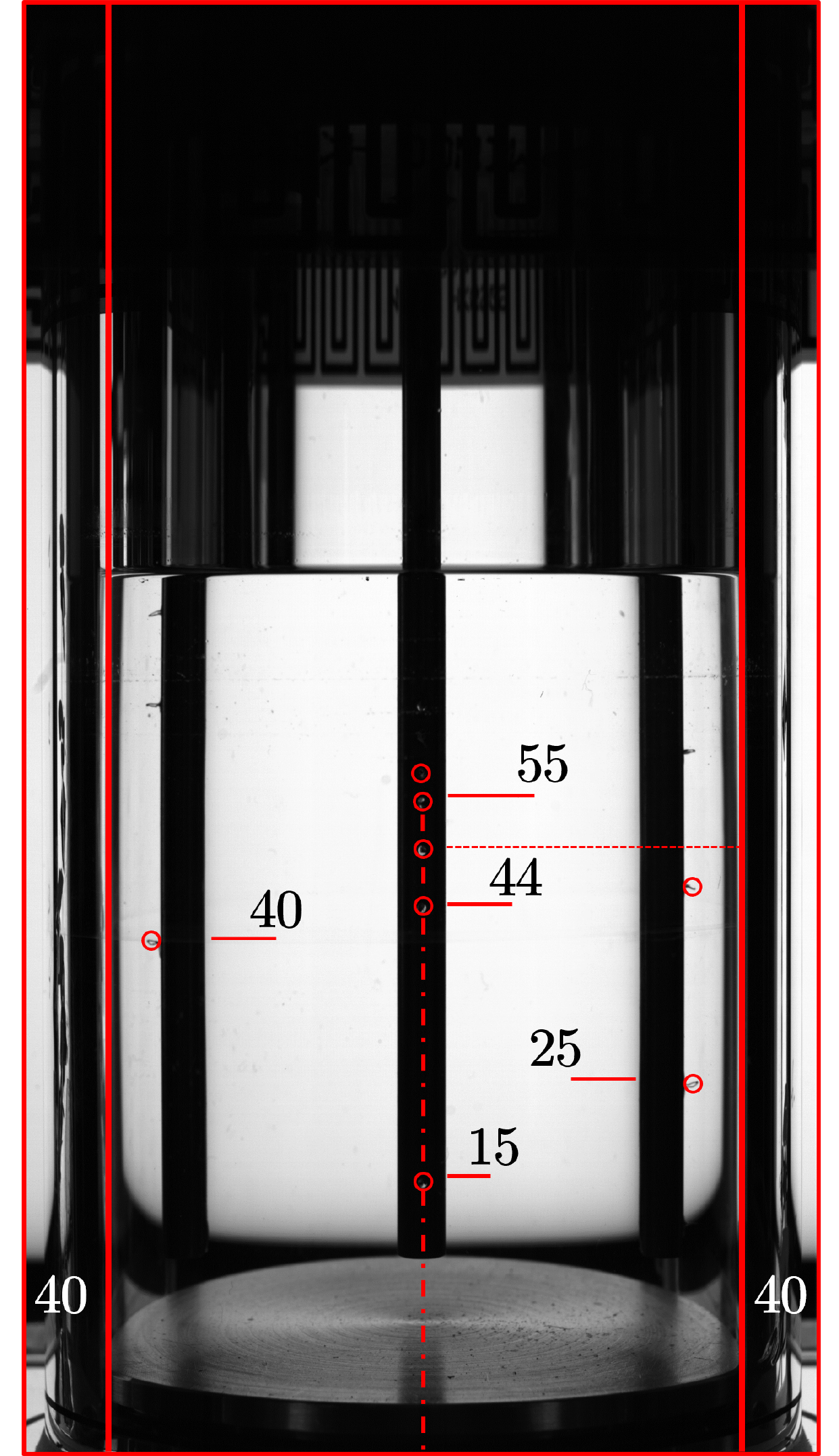}
        \caption{Non-isothermal sloshing cell $N$.}
    \end{subfigure}
    \caption{Camera view of the isothermal (a) and non-isothermal (b) test cells. Contact line position at $z \approx \SI{80}{\mm}$, the liquid is located in the lower part of the cells.} 
    \label{fig:camera_pov}
\end{figure}

High-speed video was acquired with two cameras (JAI SP-12000MCXP4) installed in custom-made aluminum supports to minimize vibrations. These cameras have a \qtyproduct{22.5 x 16.9}{\mm} sensor and were equipped with interlock objectives with a focal length of \SI{35}{\mm} to provide a field of view of \qtyproduct{170.9 x 128.3}{\mm} at a distance of approximately \SI{265.8}{\mm} from the sloshing cells. Diffusive screens (LED panels) were placed at the back of each quartz cell to perform the background lighting measurements. Two examples of raw images acquired with a f-number of $f_{\#} = 8.0$ are shown in Figure \ref{fig:camera_pov}, together with the coordinates of some of the thermocouples and rods according to the reference system introduced in Figure \ref{fig:cell}. The scaling factor for both images is $12.72 \pm 0.12\, [\text{pixels/mm}]$. The 8-bit images with a resolution of 2272 $\times$ 4096 pixels were acquired at a frame rate of $\SI{120}{\Hz}$ using the Norpix Streampix software.

The gas-liquid interface was detected via image processing using the approaches described in \citet{marques_experimental_2023} and \citet{backlight_sloshing}. This involves calculating the intensity gradient using the Sobel operator and applying thresholding to obtain a binary mask representing the free surface. Contours are detected in the mask, where a polynomial fitting is applied to obtain regression coefficients.

\subsection{Measurement chain and uncertainty} \label{subsec:exp_NIDAQ}

The measurements from all the sensors were recorded through an 8-slot USB CompactDAQ (cDAQ‑9178, National Instruments) using LabVIEW at a sampling rate of \SI{240}{\Hz}. The readings from all pressure transducers were acquired through an NI-9215 module with BNC connectors, while the signals from the 3-axis accelerometer were first conditioned with an amplifier and then recorded with an NI-9215 unit. The temperature measurements were obtained via 16-channel NI-9213 input modules with the timing mode set for high resolution. Therefore, the measurements update rate was limited to approximately 1 sample per second (S/s). The solenoid valve (SV) at the pressurant reservoir inlet was manually controlled via a NI-9481 output module that responded to a Boolean operation from the graphical LabVIEW interface. The camera acquisition was synchronized with the cDAQ data through a voltage input using a NI-9215 card and the external frame grabber (FG) pulse generator.

The pressure sensors (AMS ME780) were calibrated in-house at different temperatures, covering the entire range observed during the experiment and using a Druck DPI 610 $\pm\SI{0.025}{\percent}$ FS calibrator below atmospheric conditions and a Druck DPI601 $\pm\SI{0.05}{\percent}$ FS calibrator above atmospheric conditions. The atmospheric pressure reference during calibration was retrieved from a Druck DPI150 $\pm\SI{0.01}{\percent}$ FS. The procedure proved that the thermal shift did not impact the measurements and identified a systematic uncertainty of $\pm\SI{1.0}{kPa}$ with a 95\% confidence interval. The uncertainties were propagated across the measurement chain using the Taylor series expansion method, assuming the fully propagated uncertainty follows a symmetrical probability density function \cite{paudel_uncertainty}. 

The thermocouples type K have an uncertainty $\pm\SI{0.84}{\K}$ up to $\SI{353}{\K}$ in high-resolution sampling mode. For the triaxial accelerometer, each direction was calibrated using the gravitational acceleration ($g_0\approx\SI{9.81}{\metre\per\square\second}$) as a reference, and the uncertainty was identified to be $\pm \SI{0.1}{\metre\per\square\second}$.

\section{Experiment preparation} \label{sec:exp_preparation} 

\begin{figure*}[t!]
\centering
\includegraphics[width=0.975\textwidth]{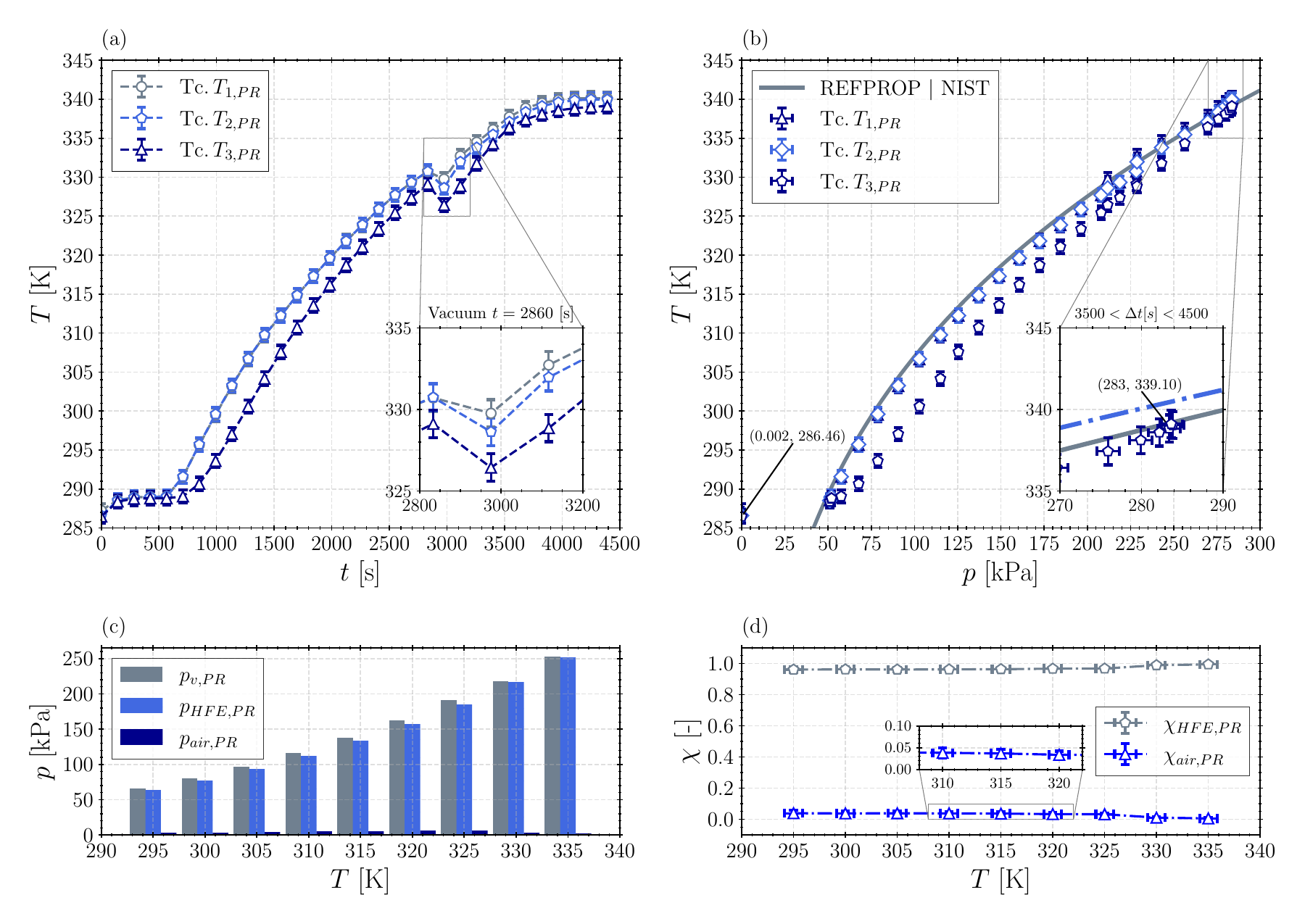}
\vspace{-5mm}
\caption{Pressurant reservoir pressure and temperature evolution during the experiment preparation phase. The time reference $t \approx 0.0$ seconds is defined when the reservoir filling starts. (a) Temperature evolution until the PID controller set-point is reached; (b) Single species analysis by comparing the reservoir pressure and temperature conditions against the saturation properties provided by the NIST REFPROP database; (c) Partial pressures and (d) mole fraction evolution during the reservoir warm-up for distinct temperatures.} 
\label{fig:single_species_PR}
\end{figure*} 

\begin{table}[b!]
\centering
\caption{Sequence of tasks for the experiment preparation.}
\renewcommand{\arraystretch}{1.025} 
\begin{tabularx}{\columnwidth}{X| *{3}{>{\centering\arraybackslash}X}} 
\toprule
\begin{tabular}[c]{@{}l@{}}Task \\ number\end{tabular} & \begin{tabular}[c]{@{}c@{}}Before takeoff  \\ ($\approx$ 2h15 min.)\end{tabular} & \begin{tabular}[c]{@{}c@{}}During steady flight  \\ ($\approx$ 20 min.)\end{tabular} \\
\midrule[1.25pt] 
1                                 & $N$ \& $I$ vacuum                                                         & -                                                                         \\                                                                
2                                & $PR$ vacuum                                                          & -                                                                         \\
3                                 & $N$ \& $I$ filling                                                         & -                                                                         \\
4                                 & $PR$ filling                                                             & $t\approx\SI{0.0}{\s}$                                                      \\
5                                 & $PR$ preheating                                                          & -                                                                         \\
6                                 & Line preheating                                                        &                                                                           \\
7                                 & $N$ cooldown                                                           & -                                                                         \\ \midrule[0.5pt] 
8                                & -                                                                     & $PR$ heating                                                                \\
9                              & -                                                                      & $N$ cooldown                                                              \\
10                               & -                                                                      & $N$ heating                                                               \\
11                               & -                                                                      & Line heating                                                              \\
12                              & -                                                                      & Acquisition ON                                                            \\
\bottomrule
\end{tabularx}
\label{tab_TASK}
\end{table}

This section outlines the experimental procedures carried out in three phases: (1) on the ground at the Novespace premises, (2) on the aircraft before takeoff, and (3) during the flight. Phase (1) begins one day before the flight and primarily focuses on preparing the single-species environment in tanks $I$, $N$, and $PR$. This phase is particularly challenging because HFE-7000 is a highly soluble solvent with an air solubility of approximately \SI{31}{\percent} in volume under normal conditions \cite{noauthor_3m_2022}. Therefore, achieving single-species conditions requires advanced degassing techniques. In this work, the de-gassification was carried out using the Freeze-pump-thaw cycling technique \cite{degassing_FPT}. This technique consists of alternating cycles of freezing, vacuuming (pumping), and thawing. These cycles were repeated three times the day before the flight in which solidification was achieved by bringing the liquid, stored in multiple borosilicate glass schlenk flasks, below $\SI{151}{\kelvin}$ (solidification point of HFE-7000 at standard atmospheric pressure) using liquid nitrogen (LN$_2$). 

\begin{figure*}[t!]
    \centering   
    \begin{subfigure}{1.0\textwidth}
        \centering
        \includegraphics[width=1.0\textwidth]{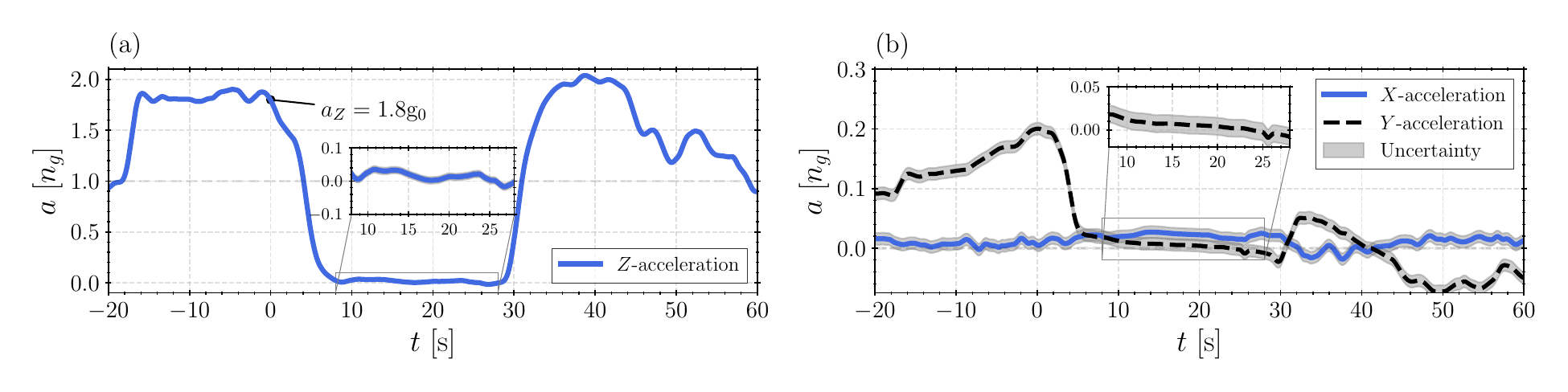}
    \end{subfigure}
    \hfill
    \begin{subfigure}{1.0\textwidth}
        \centering
        \includegraphics[width=0.57\textwidth]{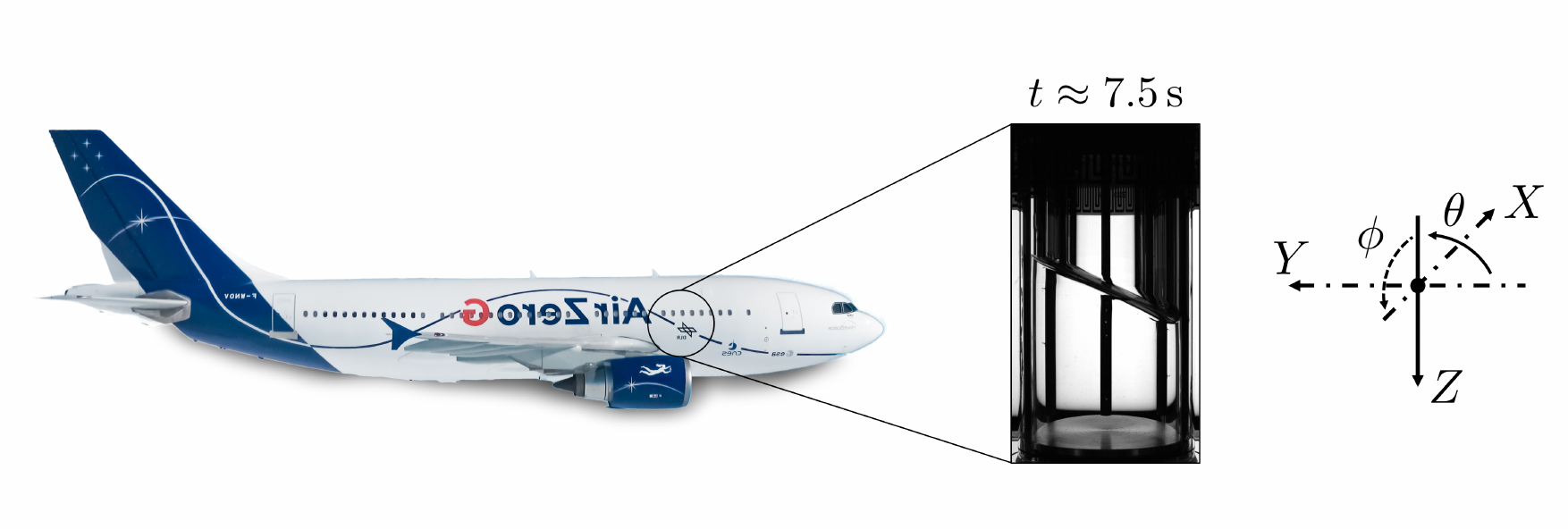}
    \end{subfigure}
\caption{Acceleration profiles for the selected experiment: parabola number 16 from the first flight $F_1(P_{16})$. The time reference $t \approx 0.0$ seconds is defined when the vertical acceleration crosses $1.8g_0$ before microgravity. (a) $Z$-acceleration and (b) $X$- and $Y$-accelerations. (Bottom): Schematic representation of the sloshing cell and the Airbus A310 with the $X$-axis aligned with the airplane's lateral axis (pitch) and $Y$-axis aligned with the longitudinal axis (roll).} 
\label{fig:accel_plane}
\end{figure*}

Phase (2), carried out 2 hours and 15 minutes before the flight, consists of introducing the liquid into the $I$, $N$, and $PR$ tanks and bringing the system to the operating conditions. This first required vacuuming all tanks and connecting lines. Table \ref{tab_TASK} reports the full sequence of tasks. Once the liquid was introduced, the $PR$ tank was heated to its operational temperature. During this phase, the quality of degassing was monitored by ensuring that the pressure and temperatures matched the expected values at saturation conditions, as provided by the NIST Reference Fluid Thermodynamic and Transport Properties Database (REFPROP, \cite{refprop}). Tasks 1-2 were performed sequentially, while tasks 3-4 were done simultaneously, maximizing the duration of the following preheating/cooling stage (tasks 6-7).

Figure \ref{fig:single_species_PR}a displays the $PR$ tank temperature evolution during the filling and heating phases, where $t\approx \SI{0.0}{\s}$ defines the beginning of task number $4$. Additionally, the temperature-pressure $p_{v, PR}$ measurements are showcased in Figure \ref{fig:single_species_PR}b for the thermocouples $T_{1, PR}$, $T_{2, PR}$, $T_{3, PR}$ (cf. Table \ref{tab:PRintrumentation}). Thermocouples $T_{1, PR}$ and $T_{2, PR}$ are in the vapor, with nearly indistinguishable readings, while Tc. $T_{3, PR}$ is in the liquid phase. The thermodynamic conditions are compared with the saturation curve obtained via the NIST REFPROP database \cite{refprop}. The measurements in the vapor phase show a remarkable agreement confirming that the system was in single-species conditions, and thus, the degassing technique was successful.

It is worth noticing that the initial temperature rise at $t<\SI{100}{\s}$ is primarily due to the compression of the residual air and liquid boiling at the onset of the filling procedure. This pressure jump is underlined in Figure \ref{fig:single_species_PR}b with the initial point before filling displayed and the coordinates highlighted for the pump base pressure ($\SI{0.002}{\kPa}$; $286.46\pm\SI{0.84}{\kelvin}$). After the filling, the reservoir stabilizes towards quasi-thermal equilibrium such that the saturation conditions could be verified at ($55\pm\SI{1.0}{\kPa}$; $288.97\pm\SI{0.84}{\kelvin}$).  

The system is kept in this condition between $\qtylist{100;700}{\s}$, after which the heating (task 5) begins, and the temperature rises towards the set point of \SI{343.0}{\K}. The heating phase takes approximately $ \SI{4000}{\s}$. Interestingly, the whole process is slow enough to ensure quasi-equilibrium conditions in the vapor phase, and the thermodynamic evolution of the system remains remarkably close to the saturation conditions (see Figure \ref{fig:single_species_PR}b) until reaching $284\pm\SI{1.0}{\kPa}$ and $340.06 \pm \SI{0.84}{\kelvin}$. On the other hand, a stratification remains in the liquid (Tc. $T_{3, PR}$), which is slightly subcooled with respect to the saturation condition at the gas-liquid interface.

Throughout this phase, at $t \approx \SI{2860}{\s}$ (see Figure \ref{fig:single_species_PR}a inset), when the temperature reached approximately \SI{330.0}{\K}, an additional vacuum purge was performed. Considering that the amount of gas dissolved in the liquid is inversely proportional to its temperature, the vacuum during this heating phase ensures the extraction of residual portions of dissolved gasses. This purging shifted the temperature measurements in Figure \ref{fig:single_species_PR}b nearer to the HFE-7000 saturation curve.

To quantify the discrepancy between the measured saturation conditions and the NIST dataset, under the assumption that the ullage volume consists of a mixture of pure vapor of HFE-7000 and air, we can estimate the partial pressure of remaining air as $p_{air, PR} = p_{v, PR}-p_{HFE, PR}$, with $p_{HFE, PR}$ provided by the REFPROP database at the measured temperature of Tc. $T_{2, PR}$. The contribution of the two species is shown in Figure \ref{fig:single_species_PR}c. Under this definition, the residual air molar fraction $\chi_{air, PR} =  1 - p_{HFE,PR}/p_{v,PR}$ could be estimated at about $0.60\pm\SI{0.54}{\percent}$. This parameter is shown in Figure \ref{fig:single_species_PR}d as a function of the measured temperature. 

The exact procedure was followed to condition the $N$ and $I$ tanks, excluding the heating and additional vacuuming at higher temperatures. As a result, we estimate a residual air molar fraction of $6.75\pm\SI{2.84}{\percent}$ for the $N$ cell and $2.88\pm\SI{1.91}{\percent}$ for the $I$ cell.

\section{Experiment conditions} \label{sec:exp_conditions} 

\begin{figure*}[t!]
  \centering   
  \begin{overpic}[width=0.95\textwidth]{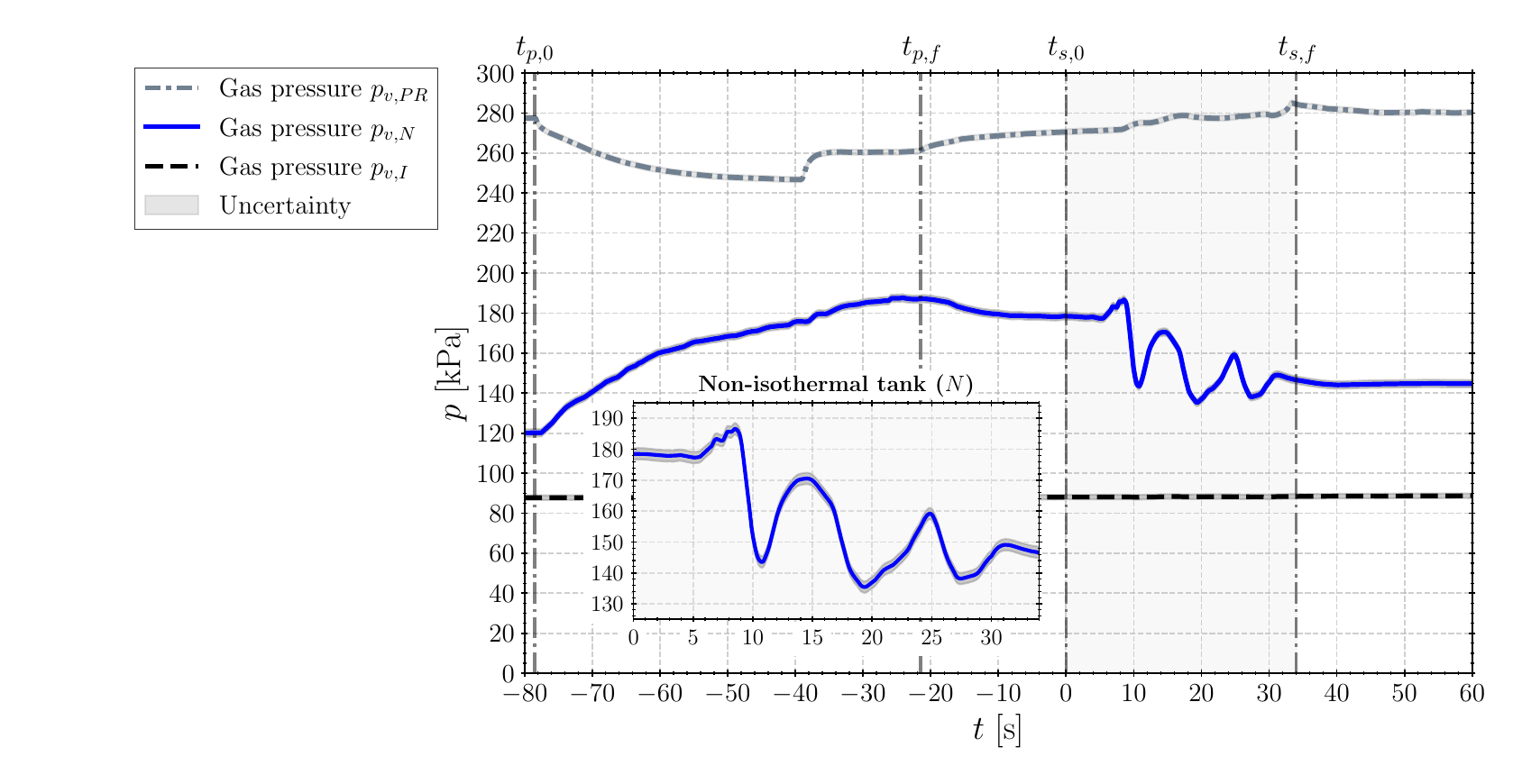}
     \put(2,3){\includegraphics[width=0.225\textwidth]{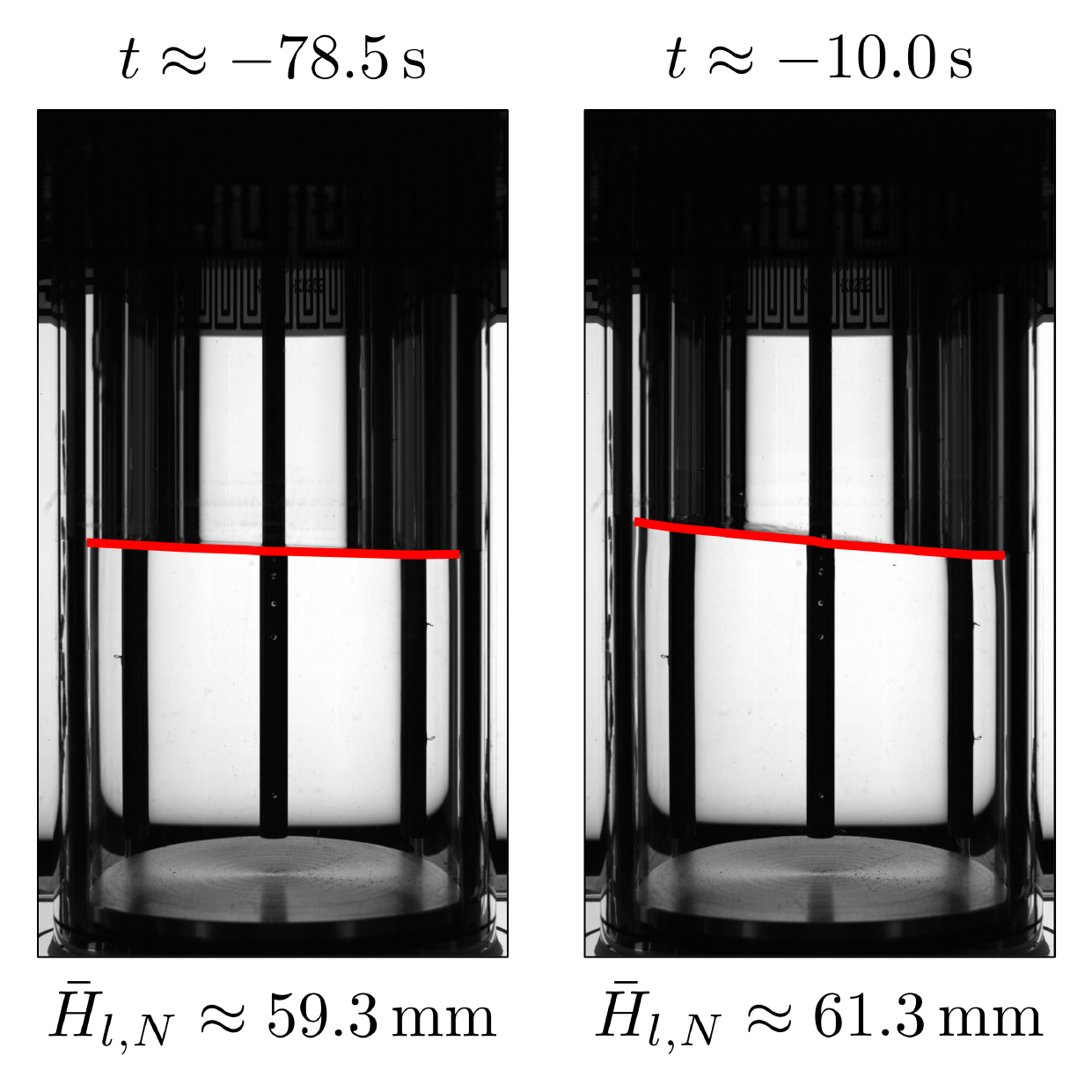}} 
  \end{overpic}
\caption{Test case $F_1(P_{16})$ absolute ullage pressure evolution for the sloshing tanks $N$ and $I$, and reservoir $PR$ as a function of time with the reference $t_{s,0}=0.0$ seconds defined when the vertical acceleration crosses $1.8g_0$ before microgravity conditions. Pressurization is set between $t_{p,0}$ and $t_{p,f}$. (Bottom left): raw images from the non-isothermal sloshing cell before and after the pressurization for liquid-level identification.} 
\label{fig:pressure_history}
\end{figure*}

\begin{table*}[t!]
\caption{Summary of the experimental conditions at $t_{s,0}=\SI{0.0}{\s}$ for different experiments at $\bar{H}_{l,N}/R \approx 1.1$, $\bar{H}_{l,N}/R \approx 1.5$ and $\bar{H}_{l,N}/R \approx 2.0$. Parabola number $16$ ($P_{16}$) is considered for the three flights. Test cases $F_1(P_{16})$ recorded during the first flight, $F_2(P_{16})$ with the lowest fill level recorded during the second flight, and $F_3(P_{16})$ in the last day at the highest fill level. The average fill level $\bar{H}_{l, N}(t_{s,0})$ for the test case with the lowest liquid volume $F_2(P_{16})$ is approximated through the readings of the thermocouples.}
\centering
\renewcommand{\arraystretch}{1.05} 
\begin{tabularx}{\textwidth}{X| *{9}{>{\centering\arraybackslash}X}} 
\midrule[1.25pt] 
Case &  Fill ratio $\bar{H}_{l,N}/R$ [-] & $\bar{H}_{l,N}(t_{s,0})$ [mm]  & $p_{v,N}(t_{s,0})$ [kPa] & $T_{sat,N}(t_{s,0})$ [K] & $T_{l,N}(t_{s,0})$ [K] & $\Delta T_{w,N}(t_{s,0})$ [K] & $\Delta T_{sub,N}(t_{s,0})$ [K] & $\Delta T_{sup,N}(t_{s,0})$ [K]  \\ \hline
$F_1(P_{16})$      & $\approx 1.5$      &  61.5       & 179        & 323.93        & 302.70      & 29.49       & 21.23      & 9.65            \\

$F_2(P_{16})$      & $\approx 1.1$       & $\approx 44$       & 149        & 318.43        & 294.77      & 35.88       & 23.66      & 16.39            \\

$F_3(P_{16})$      & $\approx 2.0$        & 81.1       & 183        & 324.75        & 305.59      & 26.50       & 19.16      & 7.68            \\
\bottomrule
\end{tabularx}
\label{tab:experimental_test_matrix}
\end{table*}

The experiments were carried out on board the 83$^{\text{rd}}$ ESA parabolic flight campaign operated by Novespace, which took place from November 20$^{\text{th}}$ to December 1$^{\text{st}}$, 2023, in Bordeaux-Mérignac (Bordeaux airport).

The Novespace parabolic flights use a modified Airbus A310 aircraft that allows controlled maneuvers to create periods of microgravity. This is accomplished by performing a flight trajectory consisting of parabolic arcs, which allows alternating two phases of hypergravity with a phase of microgravity. 

Each parabola starts with the call-out ``pull up" and the first period of hypergravity ($a_Z \approx 1.8g_0$), produced as the airplane climbs with a pitch angle $\theta$ varying from \SI{0.0}{\degree} to \SI{53.0}{\degree}. This phase lasts approximately 20 seconds, followed by a call-out ``injection" announcing the gravity step reduction. During this phase, the engines are reduced to idle, and the aircraft enters a free-fall trajectory, with microgravity conditions lasting approximately 20 - 25 seconds. The aircraft's bank angle $\phi$ is minimized throughout this stage, although values of up to \SI{8.0}{\degree} may be reached intermittently. Afterward, the call-out ``pull out" announces the beginning of a second period of hypergravity as the aircraft recovers its horizontal trajectory.

Figure \ref{fig:accel_plane} provides the acceleration profiles concerning the results discussed in section \ref{sec:results} corresponding to the first flight's 16$^{\text{th}}$ parabola with a fill level $H_l \approx \SI{60}{\mm}$. It is worth noticing that the lateral and longitudinal accelerations are non-negligible during the hypergravity phase. This is essential to trigger sloshing but challenges the repeatability of the experiment. 

Compared to experiments carried out in drop towers and sounding rockets \cite{fuhrmann_dreyer_2014, schmitt_dreyer_2015}, the lateral and longitudinal accelerations allow for an investigation focusing on chaotic sloshing rather than free-surface re-orientation. Additionally, compared to the experiments on spacecraft, the shorter microgravity phase prevents achieving a capillary-dominated spherical interface as observed in the TPCE and ZBOT experiments \cite{bentz_tank_1990, kassemi_zero-boil-off_2018}.

Overall, the campaign consisted of three flights, collecting 93 parabolas. For each flight, a specific fill ratio (namely $H_l/R=1.1$, $1.5$, and $2.0$) was used along with different injection rates, producing largely different initial conditions. This article reports the results from three parabolas, as discussed in the following section (see Table \ref{tab:experimental_test_matrix}). 

\section{Results} \label{sec:results} 

\begin{table}[t!]
\centering
\renewcommand{\arraystretch}{1.025} 
\caption{Summary of the non-isothermal sloshing tank pressure evolution during pressurization, relaxation, and microgravity injection for the 16$^{\text{th}}$ parabola of the first flight ($F_1(P_{16})$).}
\begin{tabularx}{\columnwidth}{X| *{5}{>{\centering\arraybackslash}X}} 
\toprule
$t$  [s]   & Reference & $\Delta t$ [s] & $p_{v,N}$ [kPa] & $\Delta p_{v,N}$ [kPa] & $a_{Z}$ [$n_g$]\\
\midrule[1.25pt] 
-78.5 &  $t_{p,0}$    & -        & $120 $      &  -     &   1.17    \\
-21.5 & $t_{p,f}$     & 57.0     & $187 $      &  67  &   0.99    \\
0.0 & $t_{s,0}$       & 21.5     & $179$       &  8   &   1.80    \\
8.5 & -               & 8.5     &  $188$        & 9   &   $\sim10^{-3}$   \\
28.5 & -              & 20.0     & $139$       &  49  &   $\sim10^{-3}$   \\
34.0 & $t_{s,f}$      & 5.5      & $147$       &  8   &   1.80   \\
\bottomrule
\end{tabularx}
\label{tab:pressure_evolution}
\end{table}

The results are presented in three subsections, which analyze distinct phases of the experiment. To better understand the thermodynamic evolution of each subsystem, we consider in more detail one of the three parabolas, namely $F_1(P_{16})$ (see Table \ref{tab:experimental_test_matrix} for the conditions). For this experiment, Figure \ref{fig:pressure_history} provides the gas pressure evolution in the reservoir $p_{v, PR}$, the non-isothermal sloshing cell $p_{v, N}$, and the isothermal cell $p_{v, I}$. Three phases can be identified: pressurization (from $t_{p,0}$ to $t_{p,f}$), pressure relaxation and thermal stratification (from $t_{p,f}$ to $t_{s,0}$) and sloshing (from $t_{s,0}$ to $t_{s,f}$) due to a gravity step reduction at $t_{s,0} = \SI{0.0}{\s}$. The timing $t = \SI{0.0}{\s}$ corresponds approximately to the airplane call-out ``\textit{injection}". Table \ref{tab:pressure_evolution} summarizes each phase's timing, pressure, and vertical acceleration. Further details on the temperature and liquid dynamics are described in the subsections \ref{subsec:res1} and \ref{subsec:res2}.

It is worth noting that the initial pressure in the $N$ tank $p_{v, N} \approx \SI{120}{\kPa}$ is slightly higher than in the $I$ cell due to previous active and self-pressurizations (15 parabolas). Nonetheless, the pressurization still results in a pressure ramp of $\Delta p_{v, N} \approx \SI{67}{\kPa}$ with an average rise of $\Delta p_{v, N}/\Delta t \approx \SI{1.2}{\kPa\per\s}$. Such a low flow rate ($\dot{m}_{pg}$) allows the pressurant gas to adapt to the ullage wall temperature while minimizing the gas-liquid interface impingement.

The relaxation phase between $t_{p,f} \approx \SI{-21.5}{\s}$ and $t_{s,0} \approx \SI{0.0}{\s}$ mainly occurs during the first hypergravity period with an average acceleration level of $1.8g_0$, during which the liquid thermally stratifies before the gravity step reduction triggers the sloshing event. The pressurization leads to significant condensation of the injected vapor, causing a slight change in the $N$ cell's fill level (see images on the bottom left of Figure \ref{fig:pressure_history}). During this stage, the pressure reaches a plateau of approximately $p_{v, N} \approx \SI{179}{\kPa}$ before microgravity conditions.

The pressure evolution throughout the sloshing phase from $t_{s,0} \approx \SI{0.0}{\s}$ to $t_{s,f} \approx \SI{34.0}{\s}$ is the main focus of this investigation. Interestingly, this quantity remains nearly constant up to $t \approx \SI{7.5}{\s}$ and features large fluctuations afterward. At the end of the sloshing event, as a result of the thermal destratification, the pressure drops significantly to $p_{v, N} \approx \SI{147}{\kPa}$. Nevertheless, due to the system's increase in thermal energy during the pressurization, the initial value of $p_{v, N} \approx \SI{120}{\kPa}$ is never reached. On the other hand, it is interesting to observe that the pressure in the $I$ tank remains steady at $p_{v, I} \approx \SI{88}{\kPa}$ (corresponding to a saturation temperature of \SI{303.59}{\kelvin}), highlighting that non-isothermal effects purely drive the fluctuations observed in the $N$ cell. 

\begin{figure}[b!]
    \centering   
    \begin{subfigure}{1.0\columnwidth}
        \centering
        \includegraphics[width=1.0\textwidth]{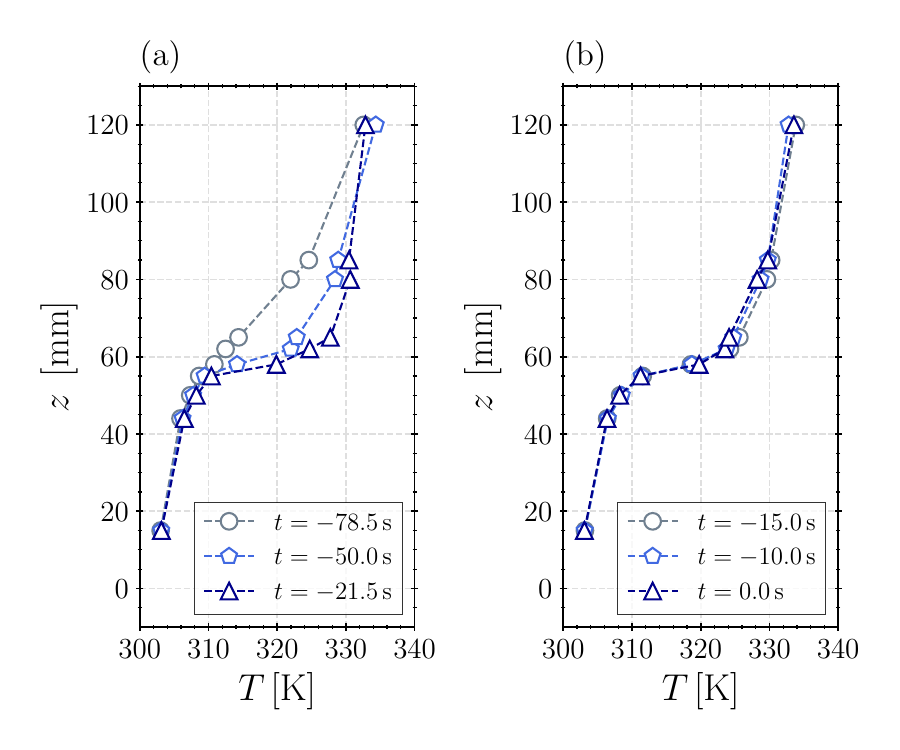}
    \end{subfigure}
    \hfill
    \begin{subfigure}{1.0\columnwidth}
        \centering
        \includegraphics[width=1.0\textwidth]{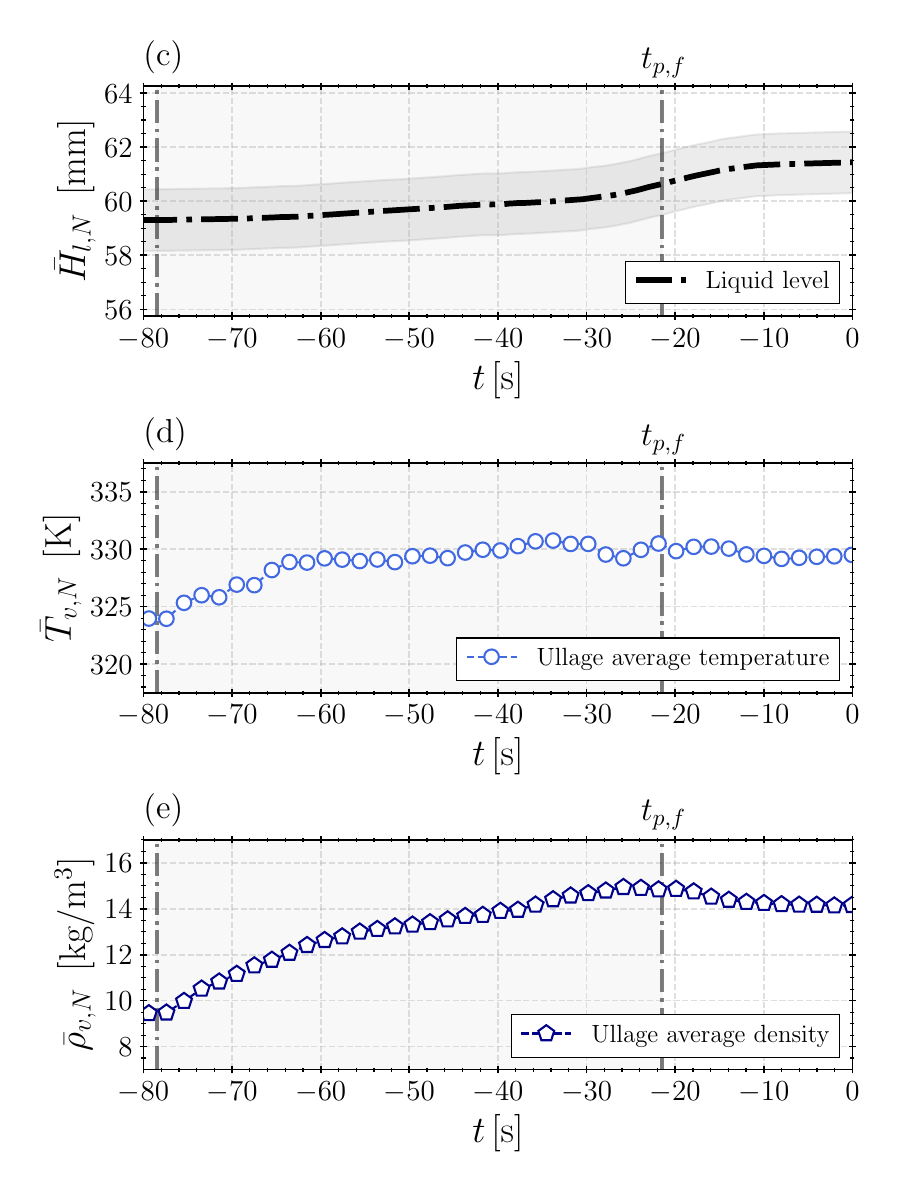}
        \end{subfigure}
  \caption{Temperature profiles and average ullage properties evolution during pressurization and relaxation $t \in [-78.5; 0.0]$ seconds in the non-isothermal sloshing cell for the test $F_1(P_{16})$. (a) Thermal stratification profiles during pressurization and (b) relaxation phases considering the temperature measurements from rod $B$ thermocouples. (c) Average liquid level $\bar{H}_{l,N}$ retrieved from the image sequences. (d) Average ullage temperature $\bar{T}_{v,N}$ and (e) density $\bar{\rho}_{v,N}$ evolution.} 
\label{fig:quantities_pressurisation}
\end{figure} 

\begin{figure*}[!h]
\centering
\includegraphics[width=1.0\textwidth]{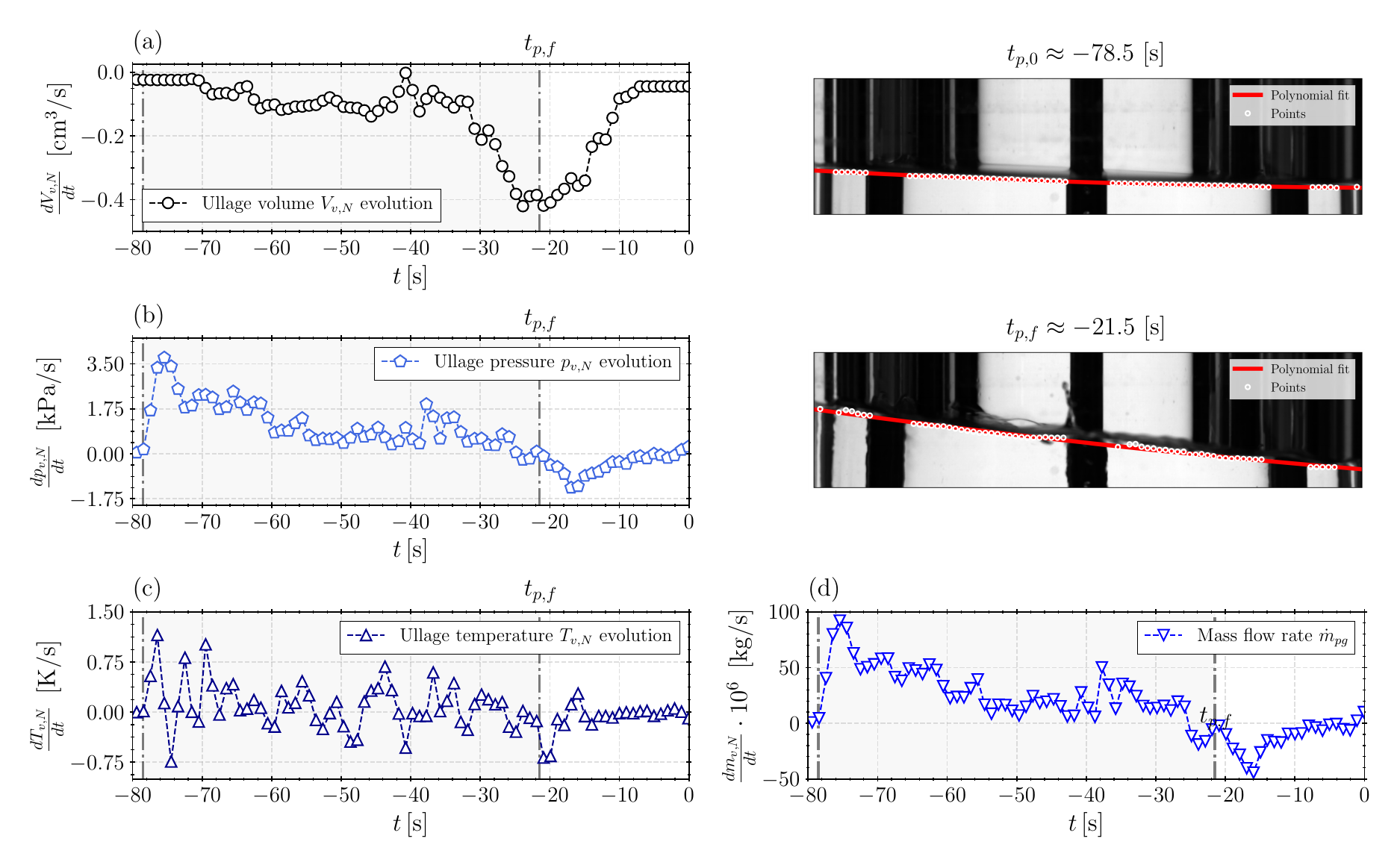}
\caption{Temporal derivatives retrieved from experimental data to determine the pressurant gas mass flow rate $\text{d}m_{v,N}/\text{d}t = \dot{m}_{pg}$ using \eqref{eq:dm_dt}. (a) Ullage volume $V_{v,N}$ time derivative determined via the liquid interface motion through the recorded image sequences. (b) Ullage pressure $p_{v,N}$ time derivative resolved from the pressure sensor measurements. (c) Average ullage temperature $\Bar{T}_{v,N}$ time derivative computed from \eqref{eq:T_u} and \ref{eq:rho_u} with the temperature measurements from the rod $B$ thermocouples exceeding the liquid interface. (Top right): post-processed image snapshots at the start ($t_{p,0}=\SI{-78.5}{\s}$) and end ($t_{p,f}=\SI{-21.5}{\s}$) of the pressurization, with a polynomial fitting resolving the interface for liquid level identification.} 
\label{fig:mass_flow_rate}
\end{figure*} 

\subsection{Pressurization and relaxation phases} \label{subsec:res1}

This section focuses on the thermodynamic response of the non-isothermal cell between $t_{p,0}$ and $t_{s,0}$. Besides providing the initial conditions for the sloshing phase, data from the pressurization phase can be used to validate numerical tools for self and active pressurizing propellant tank dynamics \cite{zimmerman_review_2013, zilliac_modeling_2005}. The temperature and pressure measurements in the cell allow for estimating the flow rate during injection based on the thermodynamic properties of the fluid. More specifically, the pressure within the cell could be linked to the instantaneous mass $m_{v,N}$, average temperature $\overline{T}_{v,N}$ and ullage volume $V_{v,N}$ in the cell:

\begin{equation}
p_{v,N}(t) = f(m_{v,N}(t), \overline{T}_{v, N}(t), V_{v,N}(t))\,,
\label{eq:equation_state}
\end{equation} where the mass averaged vapor temperature $\overline{T}_{v, N}(t)$ is defined as 

\begin{equation}
\bar{T}_{v,N}(t)=\frac{1}{\bar{\rho}_{v,N}(t)\,V_{v, N} (t)} \int_{V_{v, N}} \rho_{v,N}(z, t) \, T_{v,N}(z, t)\,\mathrm{d}V_{v, N}\,,
\label{eq:T_u}
\end{equation} assuming that both temperature and densities are axisymmetric and having introduced the average ullage density $\overline{\rho}_{v,N}$:

\begin{equation}
\bar{\rho}_{v,N}(t)=\frac{1}{V_{v, N}} \int_{V_{v, N}} \rho_{v,N}(z, t) \,\mathrm{d}V_{v, N}\,.
\label{eq:rho_u}
\end{equation}

Taking the time derivative of \eqref{eq:equation_state} and using the chain rule gives:

\begin{equation}
\frac{\text{d}p_{v,N}}{\text{d}t}=  \frac{\partial p_{v,N}}{\partial V_{v,N}}\left[\frac{\text{d} V_{v,N}}{\text{d} t}\right]+\frac{\partial p_{v,N}}{\partial m_{v,N}}\left[\frac{\text{d} m_{v,N}}{\text{d} t}\right] +\frac{\partial p_{v,N}}{\partial T_{v,N}}\left[\frac{\text{d} T_{v,N}}{\text{d} t}\right]\,,
\label{eq:dp_dt}
\end{equation} from which:

\begin{equation}
\frac{\text{d} m_{v,N}}{\text{d} t}= \Biggl(\frac{\text{d}p_{v,N}}{\text{d}t}-\frac{\partial p_{v,N}}{\partial V_{v,N}}\left[\frac{\text{d} V_{v,N}}{\text{d} t}\right]-\frac{\partial p_{v,N}}{\partial T_{v,N}}\left[\frac{\text{d} T_{v,N}}{\text{d} t}\right]\Biggr)\Bigg{/}\frac{\partial p_{v,N}}{\partial m_{v,N}}
\label{eq:dm_dt}
\end{equation}

All partial derivatives in \eqref{eq:dm_dt} can be evaluated from the fluid properties while the total derivatives are retrieved from experimental data, using a second-order finite difference scheme on the filtered signals of $p_{v,N}$, ${V_{v,N}}$ and $\overline{T}_{v,N}$. The ullage volume was computed from the liquid level observed from image acquisition. Concerning the definition of the average temperature $\overline{T}_{v,N}$ and the average density $\overline{\rho}_{v,N}$, these were computed by integrating the measurement profiles using Simpson's rule. In both calculations, the density distribution $\rho_{v,N}(z,t)$ was obtained from the (local) temperature and the (global) pressure measurement using the REFPROP database  \cite{refprop} for superheated gas conditions.

\begin{table}[t!]
\centering
\renewcommand{\arraystretch}{1.025} 
\caption{Summary of the non-isothermal sloshing cell added masses during pressurization and relaxation. The volumes are corrected for the solid elements in the tank.\\}
\begin{tabularx}{\columnwidth}{X| *{5}{>{\centering\arraybackslash}X}} 
\toprule
$t$  [s]   &  $T_{sat,N}$ [K] & $\dot{m}_{pg} \cdot 10^{6}$ [kg/s] & $\Bar{H}_{l,N}$ [mm] &$V_{v,N}$ [cm$^{3}$] & $V_{l,N}$ [cm$^{3}$]\\
\midrule[1.25pt] 
-78.5       &       312.12      &      -              &  59.3   &   315.5     &  293.6  \\
-75.5       &       314.11      &   $\approx 92.1$    &  59.4   &   315.4     &  293.7   \\
-50.0       &       322.14      &   $\approx 6.99$    &  59.7   &   313.5     &  295.6   \\
-21.5       &       325.42      &      -              &  60.6   &   308.9     &  300.2   \\
0.0         &       323.94      &       -             &  61.5   &   304.9     &  304.2   \\
\bottomrule
\end{tabularx}
\label{tab:pressurisation_masses}
\end{table}

Focusing on parabola $F_1(P_{16})$, Figure \ref{fig:quantities_pressurisation}a and \ref{fig:quantities_pressurisation}b show the measured temperature profile at five time instants, while Figure \ref{fig:quantities_pressurisation}c through \ref{fig:quantities_pressurisation}e display the time evolution of the average liquid level $\bar{H}_{l, N}$, the average ullage temperature, and density, respectively. The relevant quantities at each elected time probe are provided in Table \ref{tab:pressurisation_masses}.  Figure \ref{fig:mass_flow_rate} shows the time evolution for all the time derivatives in \eqref{eq:dm_dt} and the inferred mass flow rate together with two snapshots of the gas-liquid interface. 

\begin{figure*}[t!]
\centering
\includegraphics[width=0.8\textwidth]{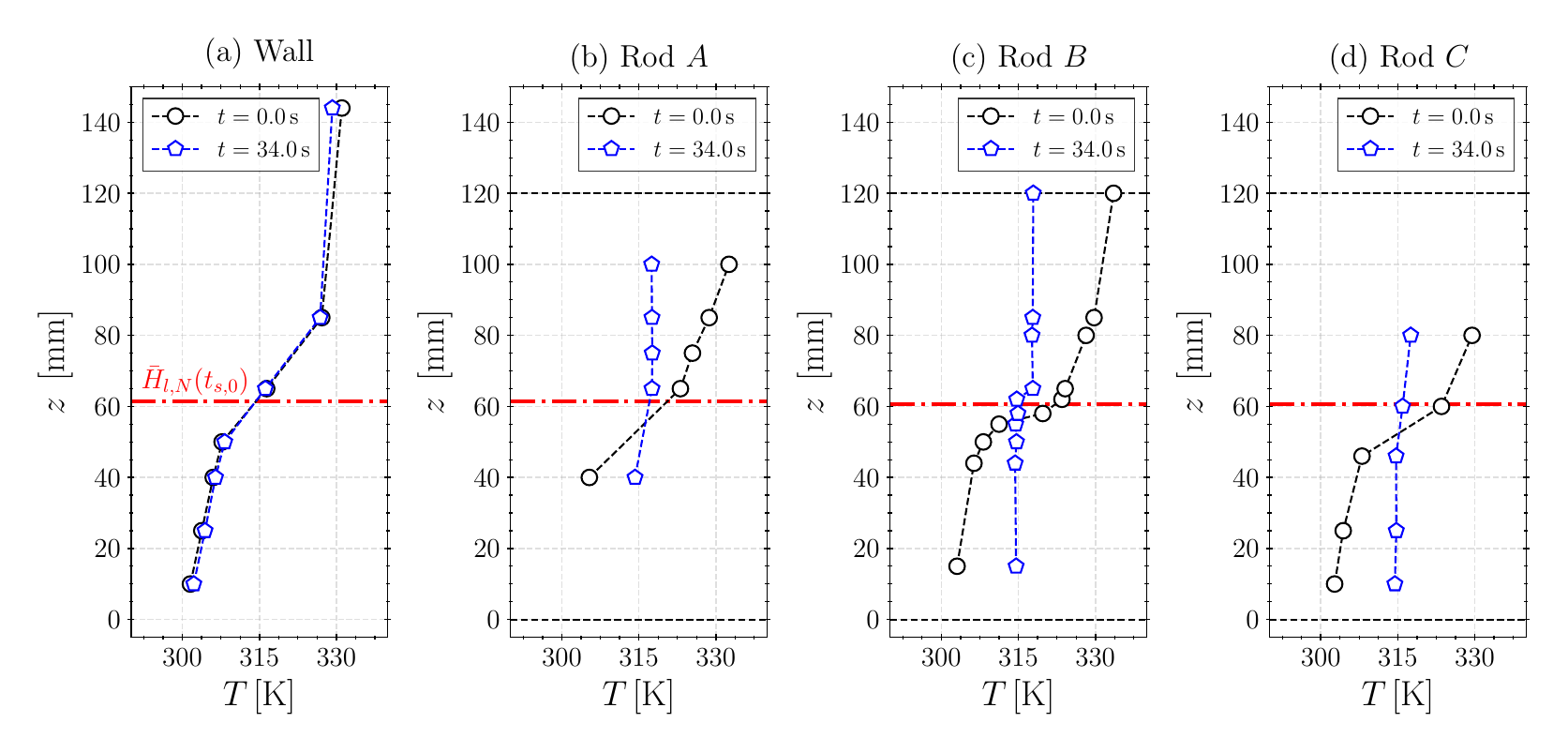}
\caption{Temperature profiles at the gravity step reduction ($t_{s,0}=\SI{0.0}{\s}$) and at the gravity step increase ($t_{s,f}=\SI{34.0}{\s}$) in the non-isothermal sloshing cell for test case $F_1(P_{16})$. (a) Quartz walls thermal stratification using the temperature measurements from reference $D$ thermocouples assuming axisymmetric profiles. (b) Temperature distribution inside the cell near the left wall is measured through the thermocouples $A_{1,N}$ to $A_{5,N}$. (c) Thermal stratification at the cell center is determined through thermocouples $B_{1,N}$ to $B_{10,N}$. (d) Temperature distribution inside the cell close to the right wall is measured through the thermocouples $C_{1,N}$ to $C_{5,N}$.} 
\label{fig:stratification_t_0.0_F1_P16}
\end{figure*}

\begin{figure*}[t!]
\centering
\includegraphics[width=0.8\textwidth]{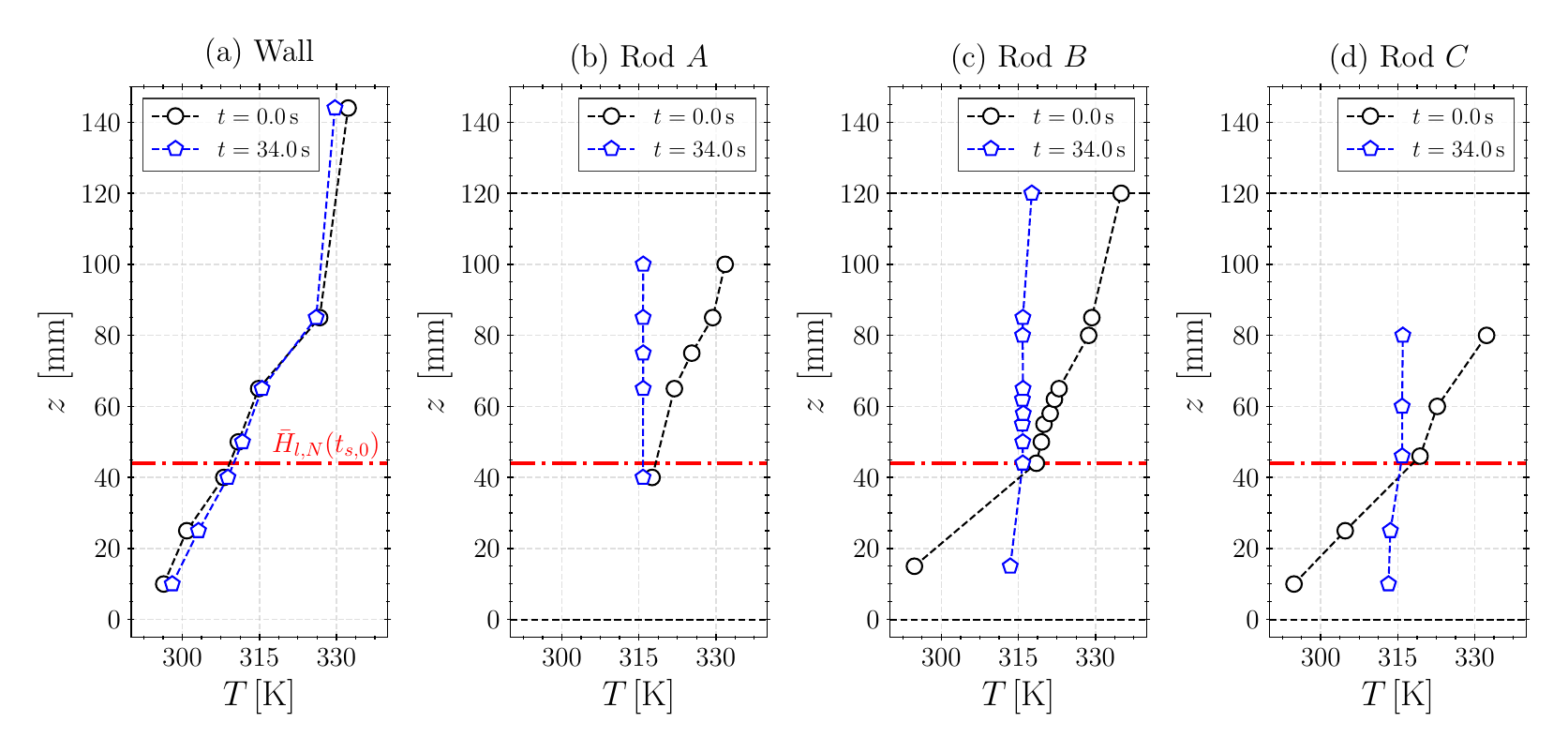}
\caption{Same as Figure \ref{fig:stratification_t_0.0_F1_P16}, but considering a parabola on the second day, with $\overline{H}_{l,N} \approx 1.1$ (see Table \ref{tab:experimental_test_matrix}).} 
\label{fig:stratification_t_0.0_F2_P16}
\end{figure*}

\begin{figure*}[t!]
\centering
\includegraphics[width=0.8\textwidth]{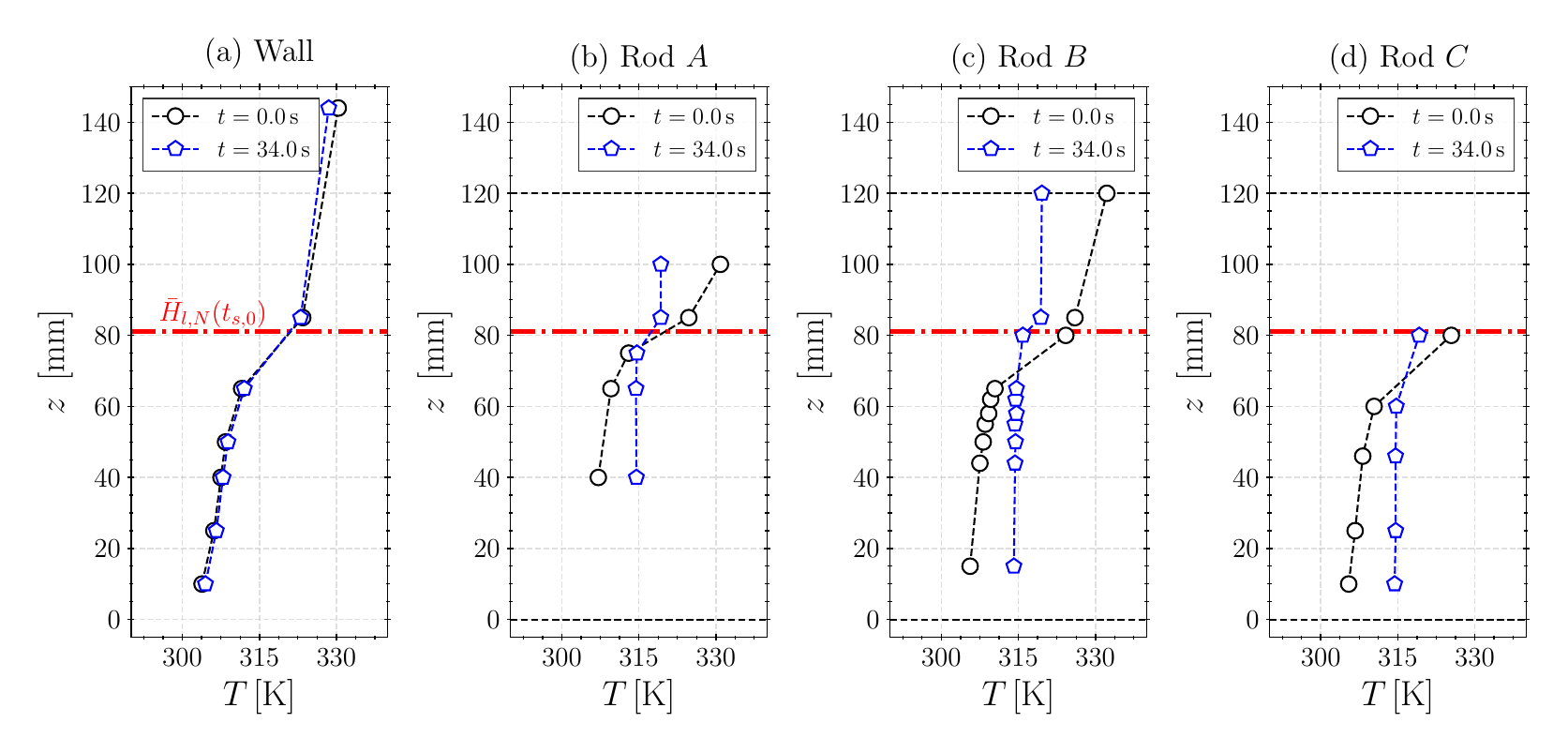}
\caption{Same as Figure \ref{fig:stratification_t_0.0_F1_P16}, but considering a parabola on the third day, with $\overline{H}_{l,N} \approx 2.0$ (see Table \ref{tab:experimental_test_matrix}).} 
\label{fig:stratification_t_0.0_F3_P16}
\end{figure*}

At the pressurization onset (Figure \ref{fig:quantities_pressurisation}a, $t_{p,0}=\SI{-78.5}{\s}$), the thermal profile exhibits a nearly linear trend in the gas phase with a $\Delta T_{sup, N} \approx \SI{20}{\kelvin}$ between $\qtylist{62;120}{\mm}$. During the interval from $t=\SI{-78.5}{\s}$ to $t=\SI{-50.0}{\s}$, the temperature distribution undergoes the most significant change near the gas-liquid interface with the average ullage temperature rising to $\bar{T}_{v, N} \approx \SI{329.41}{\kelvin}$ reaching a quasi-steady-state condition (see Figure \ref{fig:quantities_pressurisation}d). Nevertheless, the average density continues to increase throughout the entire pressurization stage. 

The temperature profile at the end of the injection (Figure \ref{fig:quantities_pressurisation}a, $t_{p,f}=\SI{-21.5}{\s}$) displays a distinct boundary layer profile on both sides of the interface, with a discontinuity of $\Delta T_{\delta_T} \approx \SI{13.4}{\kelvin}$ between $\qtylist{44;58}{\mm}$. Therefore, near the gas-liquid interface, the liquid thermal boundary layer thickness can be estimated to reach $\delta_T \sim \SI{17.0}{\mm}$. The evolution of the thermal profile is primarily driven by conduction and natural convection, as the interface remains relatively flat and steady despite residual accelerations produced during steady flight conditions. Interestingly, the liquid bulk temperature remains nearly unchanged at $T_{l, N} \approx \SI{303.15}{\kelvin}$.

Once the pressurization stops, at $t_{p,f}=\SI{-21.5}{\s}$, an appreciable rise in the liquid level is detected, as described in Figure \ref{fig:quantities_pressurisation}c and Figure \ref{fig:mass_flow_rate}a due to condensation. However, the temperature profile (see Figure \ref{fig:quantities_pressurisation}b) in both the liquid and the gas remains remarkably constant (see Figure \ref{fig:quantities_pressurisation}d). Figure \ref{fig:mass_flow_rate}a shows that the volume rate of variation reaches a maximum at the end of the pressurization and recovers to a near zero stationary condition before the sloshing event (at $t_{s,0}=\SI{0.0}{\s}$, see also Figure \ref{fig:quantities_pressurisation}c). Therefore, the pressurization analysis reveals that the cell is in stationary conditions before the gravity step reduction. This perfectly agrees with Figure \ref{fig:quantities_pressurisation}b, which displays that the temperature profiles remain unaltered during the pressure relaxation stage. 

The measured temperature profiles enable the estimation of the relative contribution of heat conduction relative to other heat transfer mechanisms between the liquid and gas phases. Assuming the two control volumes behave as semi-infinite bodies (see for example, \cite{ludwig_dreyer2013, Vanforeest_2014, ludwig_analysis_2014}), at temperature $T_{l, N} \approx \SI{303}{\kelvin}$ and $T_{v, N} \approx \SI{333}{\kelvin}$, the contact temperature at the interface would give: 

\begin{equation}
T_{i, N}=\frac{\sqrt{\rho_l c_{p,l} k_l} T_l+ \sqrt{\rho_v c_{v,v} k_v} T_v}{\sqrt{\rho_l c_{p,l} k_l}+\sqrt{\rho_v c_{v,v} k_v}} \sim \SI{304}{\kelvin}
\end{equation} which is significantly lower than $T_{i, N} \sim \SI{323}{\kelvin}$ observed in Figure \ref{fig:quantities_pressurisation}b. Moreover, under transient heat conditions between two semi-infinite bodies, the time required to reach a thermal boundary layer thickness of approximately $\delta_T \sim \SI{17}{\mm}$ on the liquid side is estimated to be an order of magnitude larger than observed in the experiments. As such, these observations suggest that the thermal equilibrium at the end of the pressure relaxation and thermal stratification period is primarily driven by natural convection.

\subsection{Thermal destratification in microgravity} \label{subsec:res2}

This subsection details the initial thermal boundary conditions governing the non-isothermal sloshing cell, along with the subsequent temperature dynamics ($t \in [0.0; 35.0]$ seconds) during the gravity step reduction after $t_{s,0} = \SI{0.0}{\s}$. Moreover, a qualitative analysis is performed on the chaotic free surface dynamics that occur when the $Z$-acceleration falls below $a_Z<0.25g_0$, promoting capillary and inertia forces to dominate the flow field over body forces \cite{dodge_2000, abramson_1981}. 

Figures \ref{fig:stratification_t_0.0_F1_P16}, \ref{fig:stratification_t_0.0_F2_P16} and \ref{fig:stratification_t_0.0_F3_P16} provide an overview of the temperature profiles in the non-isothermal cell before and after the gravity step reduction for three fill levels, that is experiments carried out in three different days (see Table \ref{tab:experimental_test_matrix}). The initial average liquid level is identified with a red continuous dashed line, and the thermal profiles at the outside quartz walls are assumed to be axisymmetric.

In all figures, the plot on the left (a) shows the temperature profiles at the wall. Notably, these profiles do not change significantly on the lateral wall. This indicates that the heat exchanged between the solid and fluid is relatively small compared to the solid's heat capacity, making a numerical simulation with a prescribed temperature at the lateral walls a reasonable simplification. This does not apply to the top cover, where the temperature variation is significant.

These figures reveal that the fill level significantly impacts sloshing induced-thermal destratification. The case with the lowest fill ratio $F_2(P_{16})$ ($\overline{H}_{l, N}/R \approx 1.1$) (see Figure \ref{fig:stratification_t_0.0_F2_P16}) displays a more considerable variation in the liquid thermal gradients than $F_3(P_{16})$ ($\overline{H}_{l, N}/R \approx 2.0$) with the highest fill ratio (see Figure \ref{fig:stratification_t_0.0_F3_P16}). As such, it pinpoints that sloshing-induced thermal mixing is more efficient at reduced fill ratios, as evidenced by the flattening of the temperature profiles in Figure \ref{fig:stratification_t_0.0_F2_P16} after the gravity step reduction event, both on the liquid and the gas phases. Given the larger liquid thermal capacity, it is evident that the increase in liquid temperature is predominantly dominated by heat exchange with the solid, particularly from the top cover, where the most substantial variation in wall temperature is measured.

\begin{figure*}[h!]
    \centering
    \includegraphics[width=1.0\textwidth]{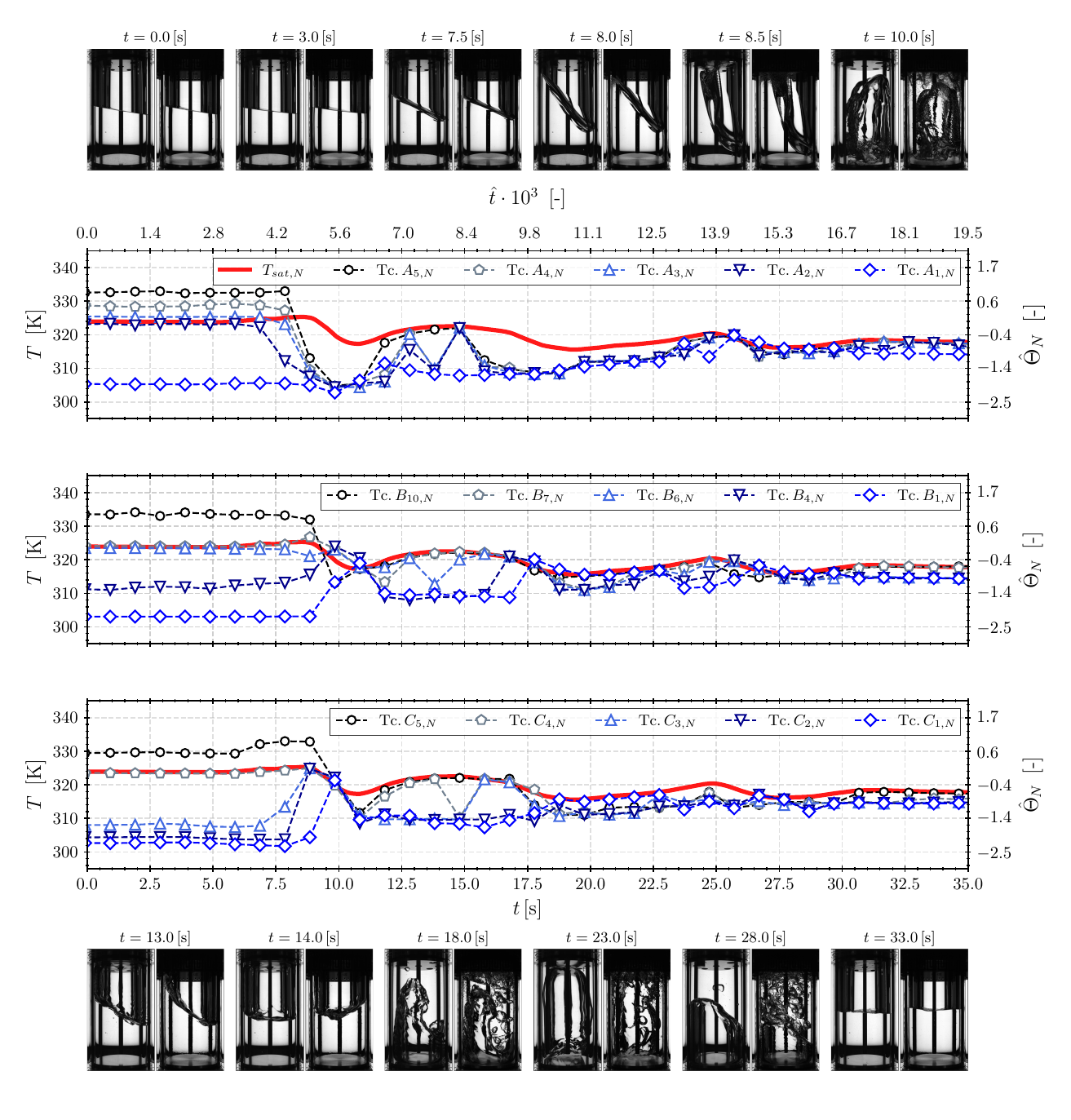}
    \caption{Temperature evolution in the different ullage and liquid thermocouples as referenced in subsection \ref{subsec:exp_NIC} for the non-isothermal sloshing cell during test case $F_1(P_{16})$. (Top): Raw images of the isothermal and non-isothermal sloshing cells during the initial gravity step reduction with the rising wave compressing the ullage vapor bubble. (a) Temperature evolution near the sloshing cell left wall for the thermocouples in rod $A$. (b) Temperature distribution for the thermocouples reference $B_{10, N}$, $B_{7, N}$, $B_{6, N}$ $B_{4, N}$ and $B_{1, N}$ located in the spatially resolved rod $B$. (c) Temperature evolution near the sloshing cell right wall for the thermocouples in rod $C$. The saturation temperature $T_{sat, N}$ is retrieved from the pressure profile, and considering the single-species assumption, it allows for tracking of the liquid and ullage phases during gravity-dominated conditions. (Bottom): Raw images of the isothermal and non-isothermal sloshing cells during the microgravity and gravity step increase phases.} 
\label{fig:temperature_F1_P16}
\end{figure*}

\begin{figure*}[h!]
    \centering
    \includegraphics[width=1.0\textwidth]{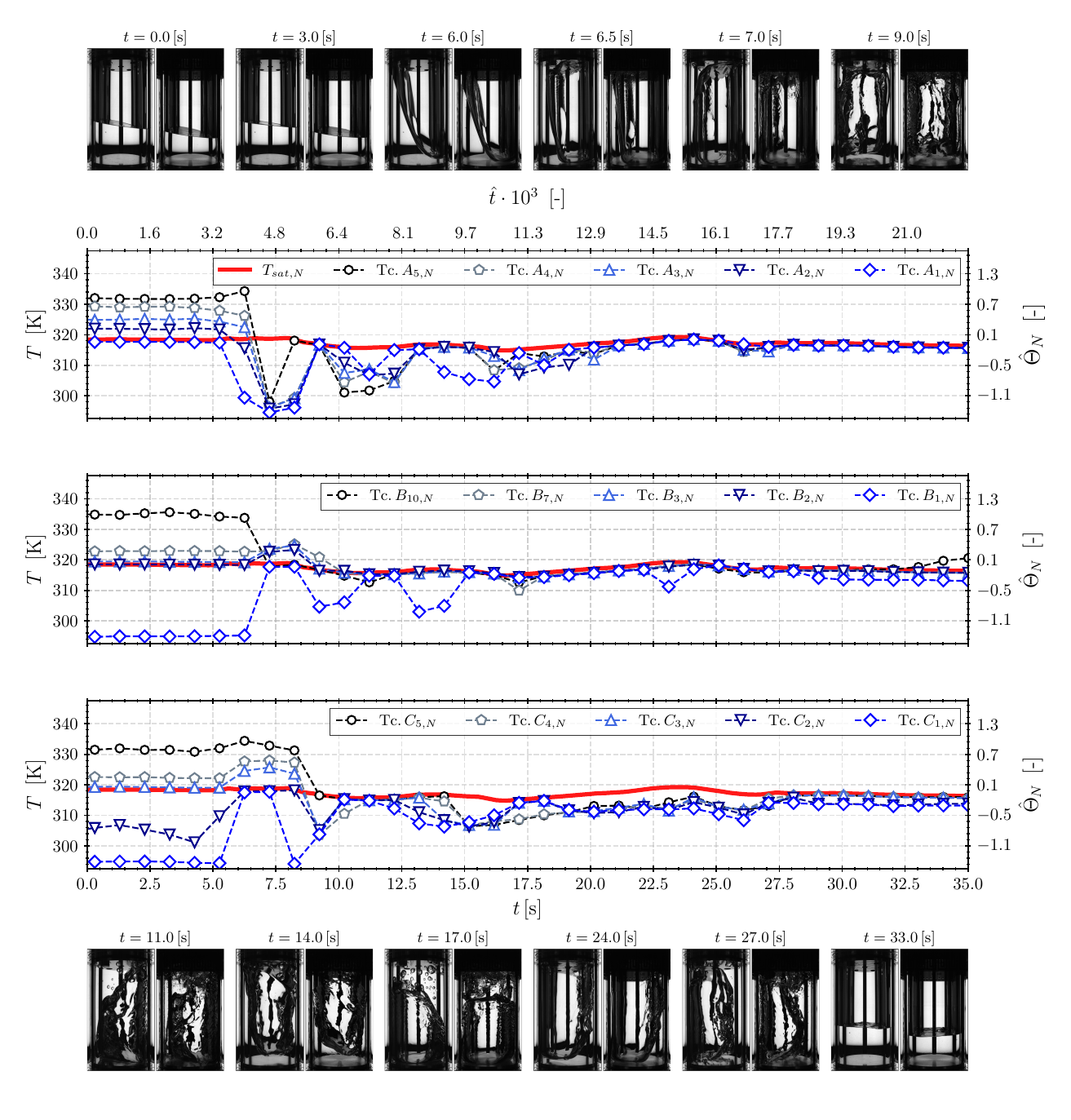}
    \caption{Same as Figure \ref{fig:temperature_F1_P16}, but considering a parabola on the second day, i.e. $F_2(P_{16})$ with $\overline{H}_{l,N} \approx 1.1$ (see Table \ref{tab:experimental_test_matrix}).} 
\label{fig:temperature_F2_P16}
\end{figure*}

\begin{figure*}[h!]
    \centering
    \includegraphics[width=1.0\textwidth]{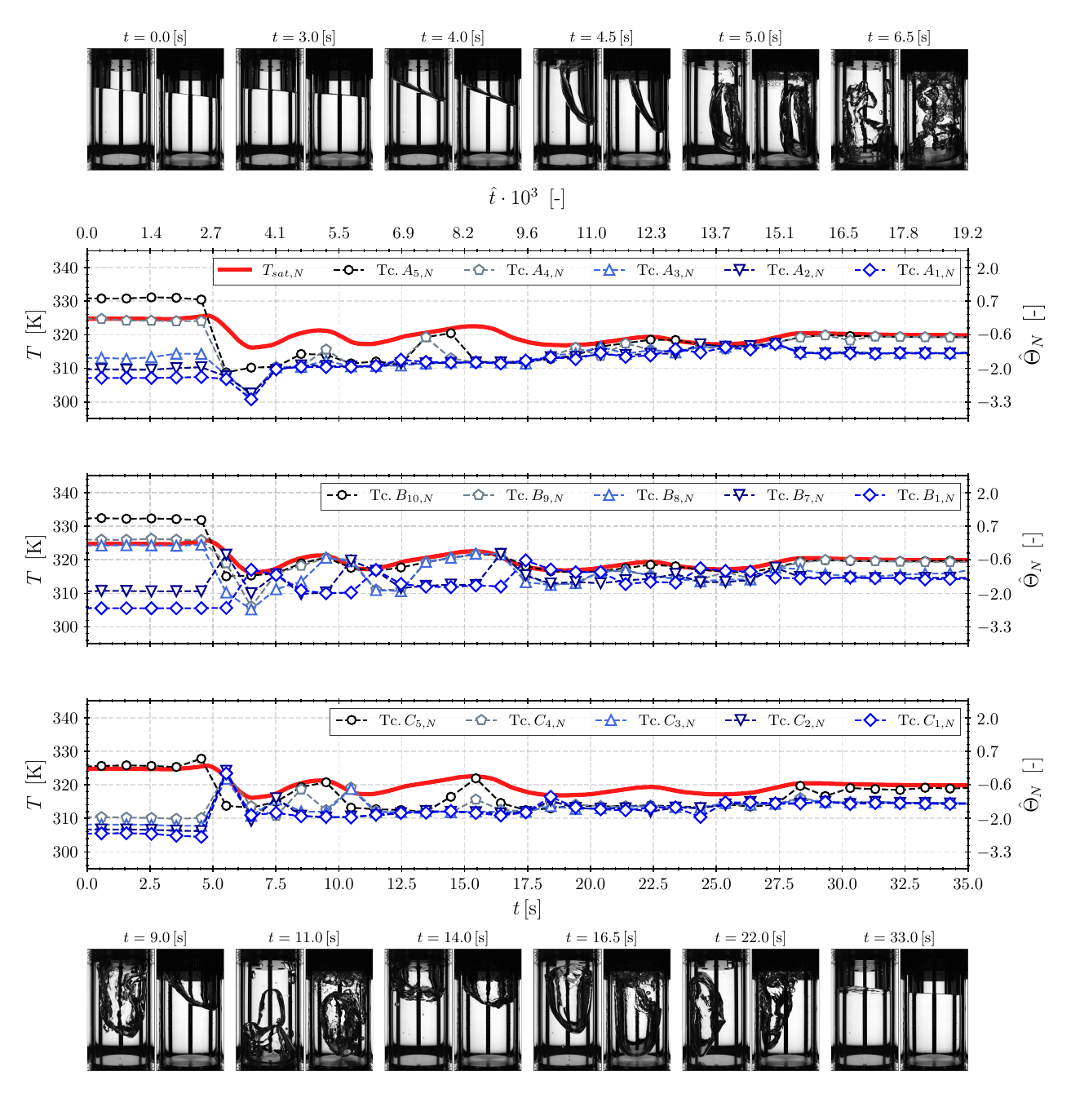}
    \caption{Same as Figure \ref{fig:temperature_F1_P16}, , but considering a parabola on the third day, i.e. $F_3(P_{16})$ with $\overline{H}_{l,N} \approx 2.0$ (see Table \ref{tab:experimental_test_matrix}).} 
\label{fig:temperature_F3_P16}
\end{figure*}

\begin{figure*}[t!]
    \centering   
    \begin{subfigure}{1.0\textwidth}
        \centering
            \includegraphics[width=0.725\textwidth]{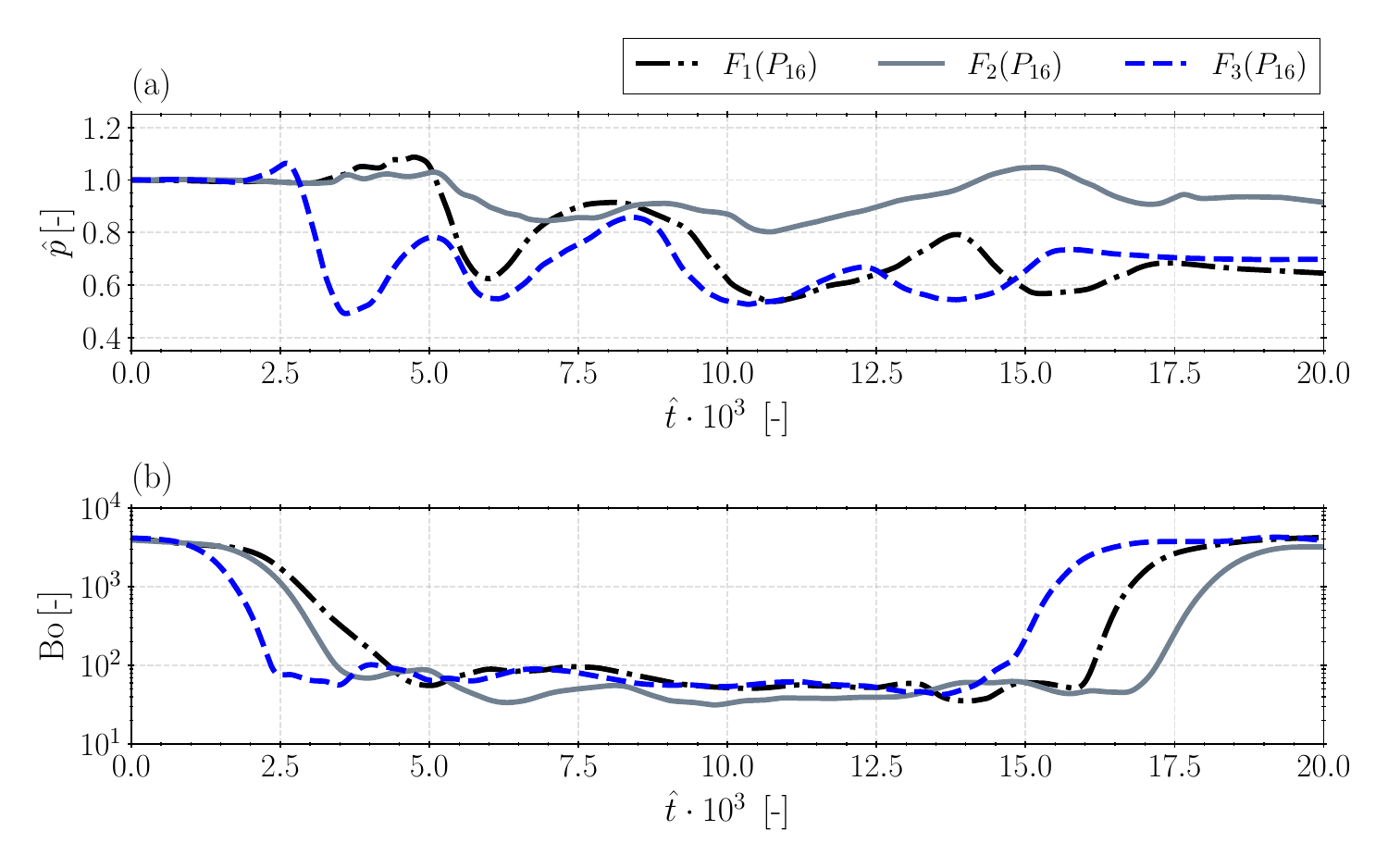}
    \end{subfigure}
  \caption{Evolution of the dimensionless pressure $(\hat{p})$ and Bond number (Bo) for the different test cases: $F_2(P_{16})$ ($\bar{H}_{l, N}/R \approx 1.1$), $F_1(P_{16})$ ($\bar{H}_{l, N}/R \approx 1.5$), and $F_3(P_{16})$ ($\bar{H}_{l, N}/R \approx 2.0$).} 
\label{fig:non-dim_pressure_BOND}
\end{figure*} 

\begin{table*}[t!]
\caption{Summary of the experimental conditions at $t_{s,0}=\SI{0.0}{\s}$ for different experiments at $\bar{H}_{l,N}/R \approx 1.1$, $\bar{H}_{l,N}/R \approx 1.5$ and $\bar{H}_{l,N}/R \approx 2.0$. Parabola number $16$ ($P_{16}$) and $21$ ($P_{21}$) are considered for the three different flights. Test cases $F_1(P_{16})$ and $F_1(P_{21})$ recorded during the first flight, $F_2(P_{16})$ and $F_2(P_{21})$ with the lowest fill level recorded during the second flight and $F_3(P_{16})$ and $F_3(P_{21})$ in the last day at the highest fill level. The average fill level $\bar{H}_{l, N}(t_{s,0})$ for the test cases with the lowest liquid volume $F_2(P_{16})$ and $F_2(P_{21})$ is approximated through the thermocouples readings.}
\centering
\renewcommand{\arraystretch}{1.025} 
\begin{tabularx}{\textwidth}{X| *{9}{>{\centering\arraybackslash}X}} 
\midrule[1.25pt] 
Case &  Fill ratio $\bar{H}_{l,N}/R$ [-] & $\bar{H}_{l,N}(t_{s,0})$ [mm]  & $p_{v,N}(t_{s,0})$ [kPa] & $T_{sat,N}(t_{s,0})$ [K] & $T_{l,N}(t_{s,0})$ [K] & $\Delta T_{w,N}(t_{s,0})$ [K] & $\Delta T_{sub,N}(t_{s,0})$ [K] & $\Delta T_{sup,N}(t_{s,0})$ [K]  \\ \hline
$F_1(P_{16})$      & $\approx 1.5$      &  61.5       & 179        & 323.93        & 302.70      & 29.49       & 21.23      & 9.65            \\
$F_1(P_{21})$      & $\approx 1.5$       & 59.7     & 177    & 323.63    & 305.88     &  27.39 &  17.75    & 10.82                 \\

$F_2(P_{16})$      & $\approx 1.1$       & $\approx 44$       & 149        & 318.43        & 294.77      & 35.88       & 23.66      & 16.39            \\
$F_2(P_{21})$      & $\approx 1.1$       & $\approx 44$       & 156        & 319.82        & 303.74      & 29.11       & 16.08      & 14.15            \\

$F_3(P_{16})$      & $\approx 2.0$        & 81.1       & 183        & 324.75        & 305.59      & 26.50       & 19.16      & 7.68            \\
$F_3(P_{21})$      & $\approx 2.0$      & 81.1       & 186         & 325.15        & 310.58      & 24.36       & 14.57      & 8.59             \\
\bottomrule
\end{tabularx}
\label{tab:experimental_test_matrix_REPEATABILITY}
\end{table*}

The analysis of the cell thermodynamic response is extended with Figures \ref{fig:temperature_F1_P16}, \ref{fig:temperature_F2_P16} and \ref{fig:temperature_F3_P16}, which provide the temperature evolution for the three test cases as a function of time, in both dimensional $T_N$ (left axis) and dimensionless $\hat{\Theta}_N$ (right axis) form for each of the thermocouples rods (Table \ref{tab:Tcs_inside_cell}). Likewise, the saturation temperature $T_{sat, N}$ is computed from the gas pressure $p_{v, N}$ and displayed in red. 

By tracking the saturation temperature $T_{sat, N}$, it is possible to identify when a given thermocouple is at the gas-liquid interface (recalling that their time response is approximately $\SI{15}{\ms}$). For example, for case $F_1(P_{16})$ in Figure \ref{fig:temperature_F1_P16}, the interface is located near thermocouples $B_{6, N}$ with $z=\SI{62}{\mm}$ at $r=\SI{0.0}{\mm}$, while at $r=\SI{33.5}{\mm}$ thermocouple $C_{4, N}$ ($z=\SI{60}{\mm}$) is the closest to interface conditions.  

For completeness, several snapshots of the liquid are also provided. In each panel, the snapshot on the left is taken from the isothermal cell, while the one on the right is retrieved from the non-isothermal tank. Animation of these three cases is provided as supplementary material to this article. The dimensionless pressure evolution $\hat{p}(\hat{t})$ and the Bond number Bo$(\hat{t})$, where the acceleration magnitude $|\bm{a}(t)|$ considers the resultant acceleration from $X-$, $Y-$, and $Z-$components, for the three test cases are shown in Figure \ref{fig:non-dim_pressure_BOND}. 

Throughout the test cases, the pressurization produces a temperature stratification on the liquid phase characterized by a sharp thermal boundary layer near the gas-liquid interface and a subcooled region with nearly homogeneous conditions underneath. As a result, the liquid's dimensionless temperature is initially negative. Depending on the parabola, the temperature remains constant during the first $5 - 7$ seconds until significant liquid motion is triggered. As the vertical acceleration falls below the threshold of $a_Z \sim 0.25g_0$, the contact line advances over the ullage walls against the flight direction, wetting it. The lateral motion is triggered by residual lateral acceleration and is consistent throughout all parabolas.

During this inertial wave, rod $A$, on the left, is initially fully covered by the liquid bulk, while rod $C$, on the right, encounters the liquid only after this has reached the top cover and encapsulated the ullage vapor (e.g., for $t > 10$ seconds for case $F_1(P_{16})$ in Figure \ref{fig:temperature_F1_P16}). The non-isothermal cell images display the formation of slim vapor bubbles near the meniscus at the heaters-ring. Therefore, boiling occurs as liquid hot spots are created at the superheated ullage walls and cover. Only at this point does a visible difference in the liquid motion between the isothermal and non-isothermal experiments become apparent. In the latter, the contact with the top cover generates boiling, which results in an inversion of the pressure evolution in cases $F_1(P_{16})$ and $F_3(P_{16})$: this is visible from the saturation temperature evolution (for example, at $t \approx \SI{8.5}{\s}$ for case $F_1(P_{16})$ in Figure \ref{fig:temperature_F1_P16} where the pressure rebuilds up to $p_{v, N} \approx \SI{188}{\kPa}$, as displayed in Figure \ref{fig:pressure_history}) and in Figure \ref{fig:non-dim_pressure_BOND}.

This is an important difference compared to on-ground experiments focusing on the destratification produced by lateral excitation (e.g., \cite{marques_experimental_2023}) where boiling is typically not produced, and the pressure evolution is dominated by condensation and evaporation at the interface. Interestingly, boiling has a significantly lower impact on the saturation temperature and, hence, pressure evolution. This is also visible in the pressure evolution in Figure \ref{fig:non-dim_pressure_BOND}. Although test case $F_3(P_{16})$ produced the most efficient mixing (as shown in Figure \ref{fig:stratification_t_0.0_F1_P16}), this is not the case producing the largest pressure drop. On the contrary, the pressure increases beyond the initial value. A possible explanation for this behavior is to be found in \eqref{eq:dp_dt}: neglecting the first term linked to the volume expansion ($\partial p_{v, N}/\partial V_{v, N}$), the second term ($\partial p_{v, N}/\partial m_{v, N}$) plays a more prominent role than the third ($\partial p_{v, N}/\partial T_{v, N}$), as temperature fluctuations do not induce significant pressure fluctuations. Moreover, for the largest ullage volume, the tank pressure is less sensitive to phase change. This could explain why the most significant pressure drop is produced for the case with the smallest ullage volume.

Overall, a remarkable agreement in liquid dynamics is visible between the isothermal and non-isothermal environments. Excluding phase-change physics, the same macro-dynamics can be retrieved during the gravity step reduction phase, with a nearly 2D inertial wave dominating the flow field. Moreover, across the test cases, it is clear that preserving the volume of both cells (as described in subsection \ref{subsec:exp_IC}) is crucial, with vapor bubbles rising to the top cover ports in the isothermal cell, which could have impacted the dynamics. Nevertheless, while the liquid is initially characterized by a single interface, as the slosh motion dampens, vapor bubbles at the $N$ tank cover start growing, splitting the liquid and creating multiple liquid-vapor interfaces where condensation/evaporation may occur. 

\subsection{A note on the test repeatability} \label{Repeatability}

\begin{figure}[b!]
    \centering
    \includegraphics[width=1.0\columnwidth]{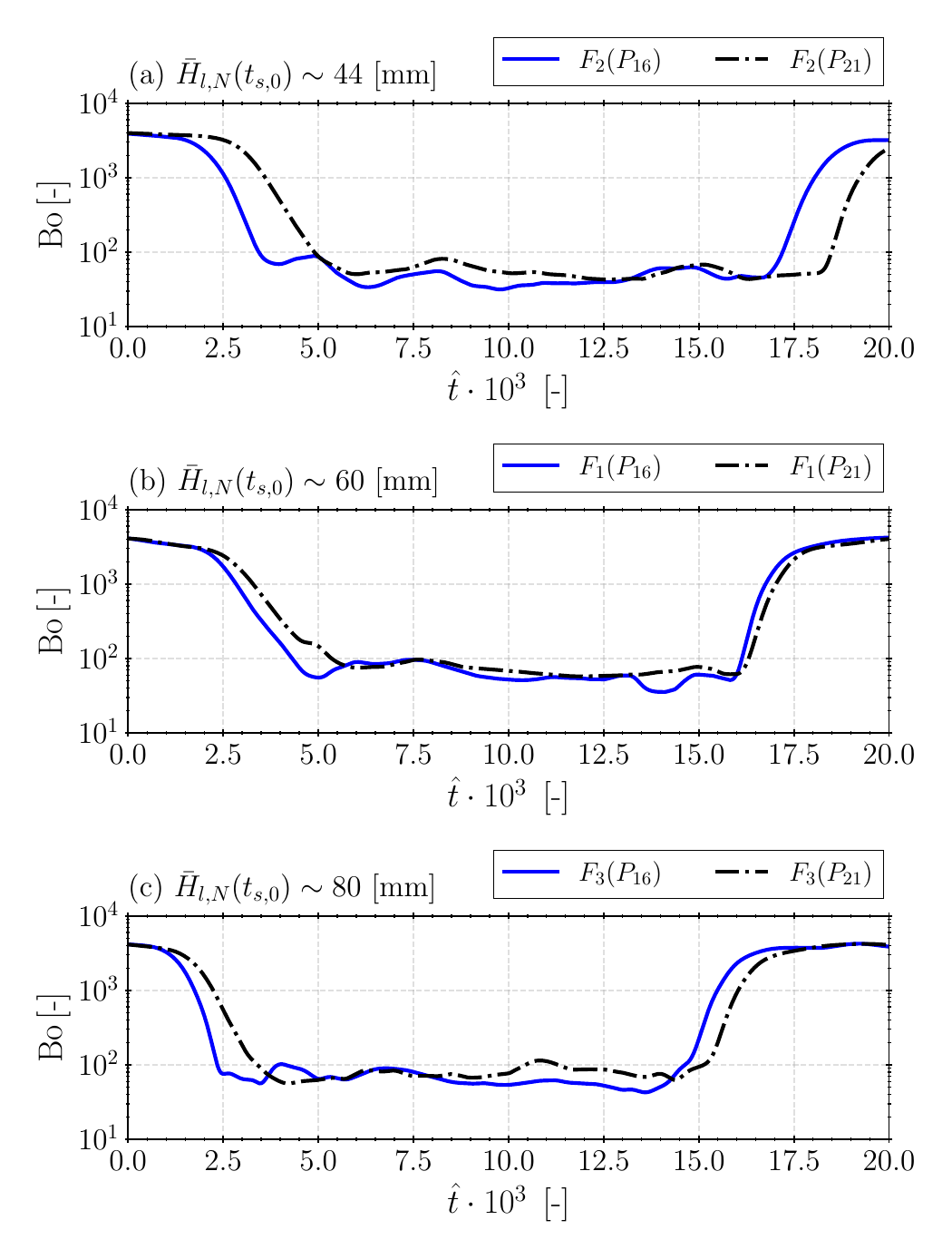}
    \caption{Evolution of the dimensionless Bond number ($\text{Bo}$) for different experiments at $\bar{H}_{l,N}/R \approx 1.1$, $\bar{H}_{l,N}/R \approx 1.5$, and $\bar{H}_{l,N}/R \approx 2.0$. The Bond number, defined as $\text{Bo}=\rho_l |\bm{a}(t)| R^2 / \sigma$ highlights the interplay between gravitational and surface tension forces in the fluid system. (a) $F_2(P_{16})$ and $F_2(P_{21})$ with the lowest fill level recorded during the second flight, (b) $F_1(P_{16})$ and $F_1(P_{21})$ recorded during the first flight at nearly \SI{50}{\percent} the tank volume, and (c) $F_3(P_{16})$ and $F_3(P_{21})$ in the last day at the highest fill level.}
    \label{fig:bond_number}
\end{figure}

The acceleration profile and the microgravity conditions achieved in each parabola depend on a series of factors, including atmospheric conditions and aircraft dynamics. Therefore, despite the sophisticated guidance, navigation, and control systems, no parabolas are alike. Moreover, the initial conditions for each parabola are different. In this brief subsection, we illustrate the impact of profile variability and differing initial conditions on the non-isothermal tank thermodynamic evolution. Table \ref{tab:experimental_test_matrix_REPEATABILITY} extends the test cases introduced in Table \ref{tab:experimental_test_matrix} considering an additional parabola each day, reporting the associated initial conditions. Figures \ref{fig:bond_number} present the dimensionless acceleration profile, expressed in terms of the Bond number, for the six cases, comparing two distinct parabolas from each day (with identical fill levels). While the differences appear minor, their impact on sloshing dynamics is significant. Videos for the additional test cases are also provided as supplementary material for this submission.

\begin{figure*}[t!]
    \centering   
    \begin{subfigure}{0.9\textwidth}
        \centering
        \includegraphics[width=0.735\textwidth]{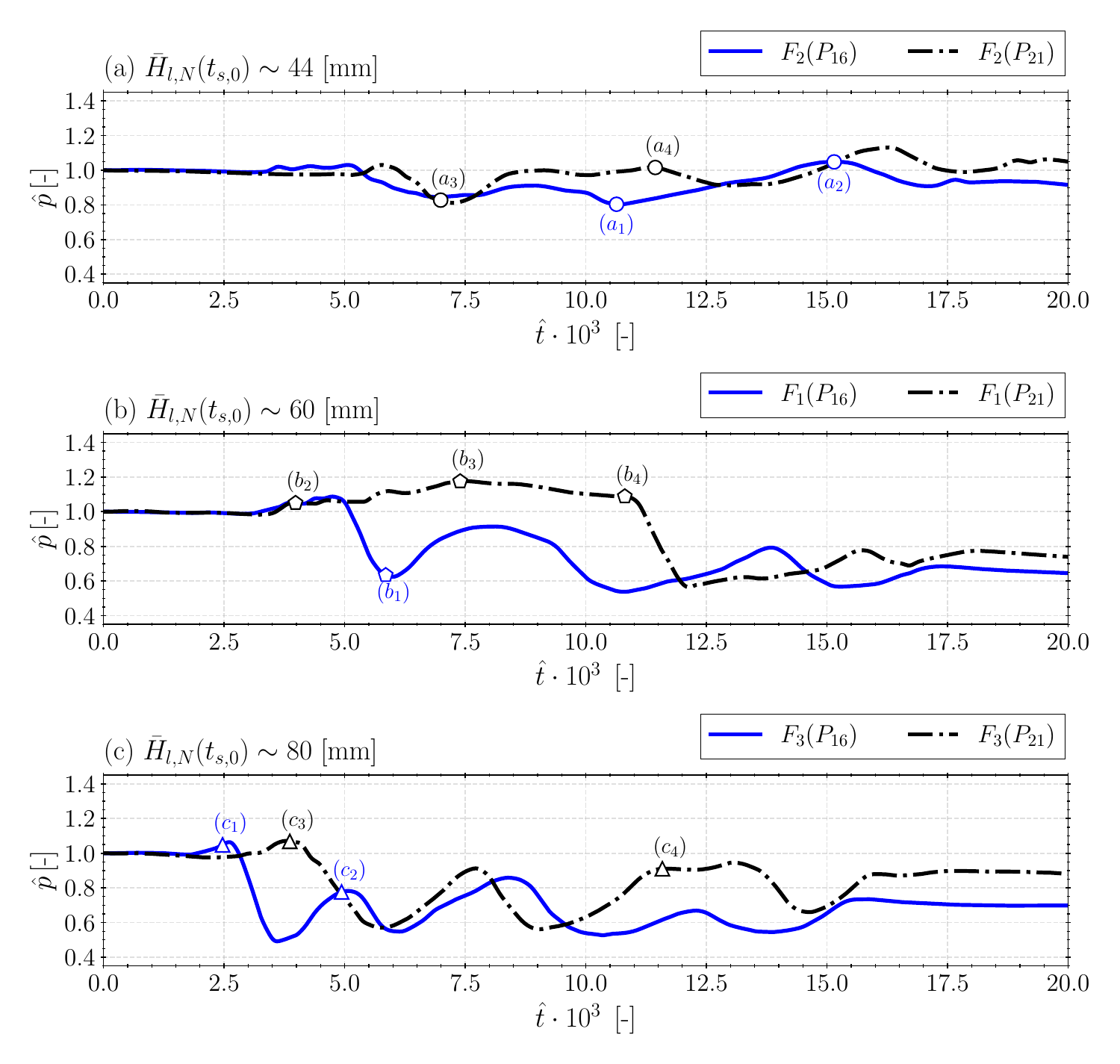}
    \end{subfigure}
    \hfill
    \begin{subfigure}{1.0\textwidth}
        \centering
        \includegraphics[width=1.0\textwidth]{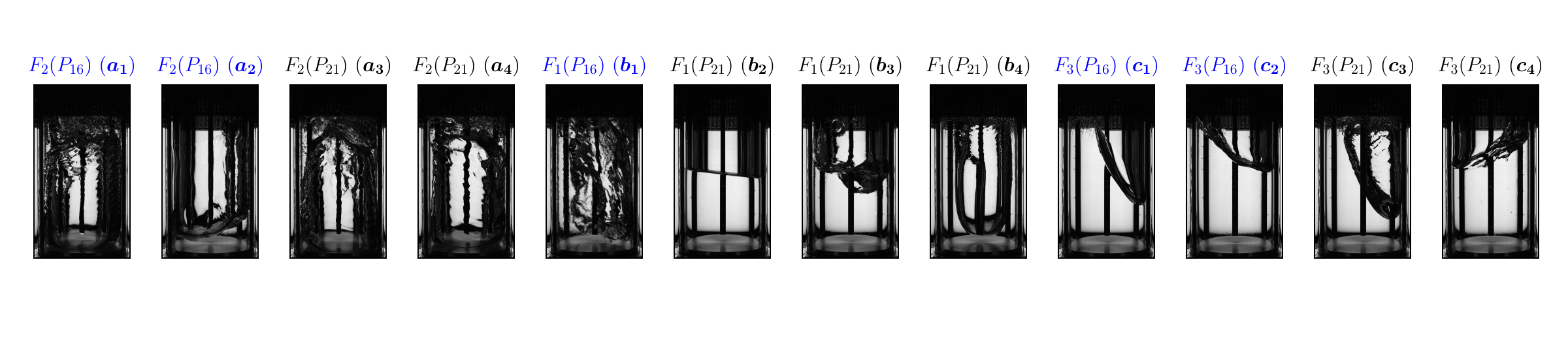}
        \end{subfigure}
  \caption{Dimensionless pressure evolution ($\hat{p}$) for different experiments at $\bar{H}_{l,N}/R \approx 1.1$, $\bar{H}_{l,N}/R \approx 1.5$, and $\bar{H}_{l,N}/R \approx 2.0$. (a) $F_2(P_{16})$ and $F_2(P_{21})$ with the lowest fill level recorded during the second flight, (b) $F_1(P_{16})$ and $F_1(P_{21})$ recorded during the first flight, and (c) $F_3(P_{16})$ and $F_3(P_{21})$ in the last day at the highest fill level. (Bottom): Raw images from the non-isothermal sloshing cell during the gravity step reduction and microgravity phases for the highlighted points from $(a_1) - (c_4)$.} 
\label{fig:non-dim_pressure}
\end{figure*} 

Minor differences in the acceleration profile result in vastly different pressure evolutions, as shown in Figure \ref{fig:non-dim_pressure}. This is particularly noticeable in the test cases from the first day:  $F_1(P_{16})$ and $F_1(P_{21})$. In $F_1(P_{21})$, the liquid interface's rise is delayed, corresponding to a delayed pressure drop. Interestingly, despite these differences in pressure evolution during the microgravity phase, the final thermodynamic states are comparable. This suggests that, while the system is highly sensitive to the acceleration profile, the overall trends in mixing-induced pressure drop, its correlation with interface dynamics, and its sensitivity to operating conditions remain consistent across all experiments.


\section{Conclusions} \label{sec:conclusions} 

In this study, we experimentally characterized the thermodynamic evolution of non-isothermal sloshing under microgravity conditions during the 83$^{\text{rd}}$ ESA parabolic flight campaign. The objective was to collect data on the evolution of temperature and pressure throughout the various phases of the sloshing event. The experimental setup consisted of a flat-bottom, upright cylindrical sloshing cell using 3M Novec 7000 as the working fluid. The facility included two identical test cells under single-species conditions: one maintained under isothermal conditions and the other subjected to an active pressurization system and non-isothermal conditions. This configuration allowed for the decoupling of dynamic effects from thermal effects, including heat and mass transfer at the gas-liquid interface.

The scaling laws of the investigated problem were introduced and used to present the results in dimensionless form. The results were analyzed for three parabolas with fill level ratios of $1.1$, $1.5$ and $2.0$. In all experiments, the tank in non-isothermal conditions was pressurized by the injection of warm vapor into its ullage up to a nominal value of approximately $\SI{180}{\kPa}$ before the gravity step reduction. Prior to the onset of sloshing, the non-isothermal cell features a stratified temperature profile, with the vapor superheated and the bulk liquid subcooled. 

The reduction in gravity caused the gas-liquid interface to rise until the tank top cover, followed by intense mixing, highlighted by chaotic behavior. No significant differences in interface dynamics were observed between the isothermal and non-isothermal cases until the liquid reached the top cover, triggering boiling in the non-isothermal cell. The pressure remained constant in the isothermal tank, whereas in the non-isothermal case, large fluctuations occurred due to phase change. Despite the pressure increase caused by boiling, the net pressure variation after sloshing is generally characterized by a drop, owing to the destratification of the initial thermal profile.

The pressure evolution within the tank is closely linked to sloshing dynamics, which are, in turn, highly sensitive to the acceleration profile. Consequently, achieving repeatability in sloshing-induced thermal destratification across parabolas with identical fill ratios proved challenging. Nevertheless, specific general trends were observed: test cases with smaller fill ratios (i.e., larger ullage volumes) produce more efficient thermal mixing but exhibit a smaller pressure drop. This is attributed to the reduced sensitivity, $\partial p/\partial m$, in the ullage volume. This work provides a unique database for validating high-fidelity simulations, and calibrating simplified thermodynamic models of propellant storage tanks, representing a first step toward developing new control strategies for minimizing boil-off losses.


\bibliographystyle{elsarticle-num-names}
\bibliography{Full_biblio} 
\addcontentsline{toc}{section}{References}

\section{Data availability} \label{sec:data_share} 

The data supporting this study's findings will be stored in a dedicated database, which is currently under development. This database will be accessible to the public upon request to the corresponding author.

\section{Acknowledgments} \label{sec:ackn} 

This work was supported by the European Space Agency (ESA) in the framework of the GSTP-SLOSHII project with reference number 4000129315/19/NL/MG and BELSPO through PRODEX fund number 4000142800 Non-isoThermal Sloshing PARabolic FliGht Experiment (NT-SPARGE). The view expressed herein cannot be taken to reflect the official opinion of the European Space Agency. The authors gratefully acknowledge the financial support of the ‘Fonds de la Recherche Scientifique (F.R.S. - FNRS)’ for the FRIA grant with reference FC47297 supporting the Ph.D. of Mr. Marques.

\appendix

\section{Nomenclature} \label{sec:appA}

\renewcommand{\nomname}{}

\vspace{-5mm}
     \nomenclature[L$u$]{$[u]$}{dimensionless velocity scale, -}
     \nomenclature[L$u$]{$[a]$}{acceleration scale, $\unit{\m\per\square\s}$}
     \nomenclature[L$cp$]{$c_p$}{isobaric specific heat, ${\unit{\joule\per\kg\per\kelvin}}$}
     \nomenclature[L$cv$]{$c_v$}{isochoric specific heat, ${\unit{\joule\per\kg\per\kelvin}}$}
     \nomenclature[L$g$]{$g_0$}{gravitational acceleration, $\unit{\m\per\square\s}$}
     \nomenclature[L$a$]{$\boldsymbol{a}$}{acceleration vector, $\unit{\m\per\square\s}$}
     \nomenclature[L$H$]{$H$}{height, ${\rm m}$}
     \nomenclature[L$p$]{$p$}{pressure, ${\unit{\pascal}}$}
     \nomenclature[L$k$]{$k$}{thermal conductivity, $\unit{\watt\per\m\per\kelvin}$}
     \nomenclature[L$L_v$]{$\mathcal{L}_v$}{latent heat of vaporization, ${\rm J/kg}$}
     \nomenclature[L$m$]{$m$}{mass, ${\rm kg}$}
     \nomenclature[L$n$]{$n$}{gas number of moles, ${\rm mol}$}
     \nomenclature[L$m_dot$]{$\dot{m}$}{mass flow, ${\unit{\kg\per\s}}$}
     \nomenclature[L$R$]{$R$}{radius, ${\rm m}$}
     \nomenclature[L$z$]{$z$}{vertical position, ${\rm m}$}
     \nomenclature[L$r$]{$r$}{radial position in cylindrical coordinates, ${\rm m}$}
     \nomenclature[L$D$]{$D$}{diameter, ${\rm m}$}
     \nomenclature[L$T$]{$T$}{temperature, ${\rm K}$}
     \nomenclature[L$t$]{$t$}{time, ${\rm s}$}
     \nomenclature[L$V$]{$V$}{volume, ${\rm m^3}$}
     \nomenclature[L$R_s$]{$\mathcal{R}_s$}{specific gas constant, $\unit{\joule\per\kg\per\kelvin}$}
     \nomenclature[L$M_w$]{$M_w$}{molar mass, ${\unit{\kg\per\mol}}$}

     \nomenclature[G$01$]{$\delta_T$}{thermal boundary layer thickness, ${\rm mm}$}
     \nomenclature[G$50$]{$\gamma$}{heat capacity ratio, -}
     \nomenclature[G$06$]{$\alpha$}{thermal diffusivity, ${\unit{\m\square\per\s}}$}
     \nomenclature[G$02$]{$\vartheta$}{angle in cylindrical coordinates, ${\unit{\degree}}$}
     \nomenclature[G$16$]{$\rho$}{mass-specific density, ${\rm kg/m^3}$}
     \nomenclature[G$17$]{$\sigma$}{surface tension, ${\rm N/m}$}
     \nomenclature[G$19$]{$\mu$}{dynamic viscosity, ${\unit{\pascal\per\s}}$}
     \nomenclature[G$20$]{$\chi$}{mole fraction, -}
     \nomenclature[G$21$]{$\theta$}{airplane pitch angle, ${\unit{\degree}}$}
     \nomenclature[G$22$]{$\phi$}{airplane bank angle, ${\unit{\degree}}$}
     \nomenclature[G$23$]{$\Theta$}{thermal stratification ratio, -}

     \nomenclature[S$^$]{$\boldsymbol{\hat{}}$}{dimensionless quantities}
     \nomenclature[S$-$]{$\boldsymbol{\Bar{}}$}{average quantities}
     \nomenclature[S$I$]{$I$}{isothermal cell}
     \nomenclature[S$N$]{$N$}{non-isothermal cell}
     \nomenclature[S$PR$]{$PR$}{pressurant reservoir}
     \nomenclature[S$v$]{$v$}{vapor}
     \nomenclature[S$l$]{$l$}{liquid}
     \nomenclature[S$i$]{$i$}{interface}
     \nomenclature[S$w$]{$w$}{walls}
     \nomenclature[S$a$]{$a$}{ambient}
     \nomenclature[S$sat$]{$sat$}{saturation}
     \nomenclature[S$s$]{$s$}{sloshing}
     \nomenclature[S$p$]{$p$}{pressurization}
     \nomenclature[S$pg$]{$pg$}{pressurant gas}
     \nomenclature[S$sub$]{$sub$}{subcooled}
     \nomenclature[S$sub$]{$sup$}{superheated}
     \nomenclature[S$c$]{$c$}{cooler}
     \nomenclature[S$h$]{$h$}{heater}

     \nomenclature[N$Bo$]{Bo}{Bond}
     \nomenclature[N$Pr$]{Pr}{Prandtl}
     \nomenclature[N$Re$]{Re}{Reynolds}
	
    \printnomenclature

\end{document}